\documentstyle[12pt]{article}
\newlength{\pubnumber} 

\catcode`\@=11
\@addtoreset{equation}{section}

\def\section{\@startsection{section}{1}{\z@}{3.5ex plus 1ex minus .2ex}
 {2.3ex plus .2ex}{\large\bf}}
\def\subsection{\@startsection{subsection}{2}{\z@}{2.3ex plus .2ex}
 {2.3ex plus .2ex}{\bf}}

\begin{document}

\begin{titlepage}
\samepage{
\rightline{CERN-TH/99-333}
\rightline{\tt hep-ph/9912455}
\rightline{December 1999}
\vfill
\smallskip
\smallskip
\begin{center}
   {\Large \bf Invisible Axions and Large-Radius Compactifications\\}
\vfill
\medskip
\smallskip
\vspace{.09in}
   {\large
    Keith R. Dienes$^{1,2}$\footnote{
     E-mail address: dienes@physics.arizona.edu},
    Emilian Dudas$^{3}$\footnote{
     E-mail address:  emilian.dudas@th.u-psud.fr},
     $\,$and$\,$
      Tony Gherghetta$^{1,4}$\footnote{
     E-mail address: tony.gherghetta@cern.ch}
    \\}
\vspace{.01in}
\vspace{.28in}
 {\it  $^1$ Theory Division, CERN, CH-1211 Geneva 23, Switzerland\\}
\vspace{.03in}
 {\it  $^2$ Department of Physics, University of Arizona,
             Tucson, AZ  85721 USA\\}
\vspace{.03in}
 {\it  $^3$ LPT, B\^at.~210, Univ.\ Paris-Sud, F-91405, Orsay Cedex, France\\}
\vspace{.03in}
 {\it  $^4$ IPT, University of Lausanne, CH-1015 Lausanne, Switzerland\\} 
\end{center}
\vfill
\begin{abstract}
  {\rm
     We study some of the novel effects that arise when the QCD axion
     is placed in the ``bulk'' of large extra spacetime dimensions.
     First, we find that the mass of the axion can become independent
     of the energy scale associated with the breaking of the Peccei-Quinn symmetry.
     This implies that the   mass of the axion can be adjusted independently of its 
     couplings to ordinary matter, thereby providing a new method of rendering
     the axion invisible.
     Second, we discuss the new phenomenon of laboratory axion oscillations 
     (analogous to neutrino oscillations), and show that these oscillations cause
     laboratory axions to ``decohere'' extremely rapidly 
     as a result of Kaluza-Klein mixing.  
     This decoherence may also be a contributing factor to axion invisibility.
     Third, we discuss the role of Kaluza-Klein axions in axion-mediated
     processes and decays, and propose several experimental tests of the 
       higher-dimensional nature of the axion. 
     Finally, we show that under certain circumstances, the presence of an infinite 
     tower of Kaluza-Klein axion modes can significantly accelerate the dissipation of
     the energy associated with cosmological relic axion oscillations, 
     thereby enabling the Peccei-Quinn symmetry-breaking scale to
     exceed the usual four-dimensional relic oscillation bounds.
     Together, these ideas therefore provide new ways of obtaining 
     an ``invisible'' axion within the context of higher-dimensional theories 
     with large-radius compactifications.
      } 
\end{abstract}
   }
\end{titlepage}

\setcounter{footnote}{0}

\newcommand{\newc}{\newcommand}

\newc{\gsim}{\lower.7ex\hbox{$\;\stackrel{\textstyle>}{\sim}\;$}}
\newc{\lsim}{\lower.7ex\hbox{$\;\stackrel{\textstyle<}{\sim}\;$}}

\def\beq{\begin{equation}}
\def\eeq{\end{equation}}
\def\beqn{\begin{eqnarray}}
\def\eeqn{\end{eqnarray}}
\def\sosixteen{{$SO(16)\times SO(16)$}}
\def\e8{{$E_8\times E_8$}}
\def\V#1{{\bf V_{#1}}}
\def\half{{\textstyle{1\over 2}}}
\def\ttwo{{\vartheta_2}}
\def\tthree{{\vartheta_3}}
\def\tfour{{\vartheta_4}}
\def\ttwob{{\overline{\vartheta}_2}}
\def\tthreeb{{\overline{\vartheta}_3}}
\def\tfourb{{\overline{\vartheta}_4}}
\def\etainv{{\overline{\eta}}}
\def\Str{{{\rm Str}\,}}
\def\bone{{\bf 1}}
\def\chibar{{\overline{\chi}}}
\def\Jbar{{\overline{J}}}
\def\qbar{{\overline{q}}}
\def\calO{{\cal O}}
\def\calE{{\cal E}}
\def\calT{{\cal T}}
\def\calM{{\cal M}}
\def\calF{{\cal F}}
\def\calY{{\cal Y}}
\def\rep#1{{\bf {#1}}}
\def\ie{{\it i.e.}\/}
\def\eg{{\it e.g.}\/}
\def\eleven{{(11)}}
\def\ten{{(10)}}
\def\nine{{(9)}}
\def\Ip{{\rm I'}}
\def\oneprime{{I$'$}}
\hyphenation{su-per-sym-met-ric non-su-per-sym-met-ric}
\hyphenation{space-time-super-sym-met-ric}
\hyphenation{mod-u-lar mod-u-lar--in-var-i-ant}


\def\inbar{\,\vrule height1.5ex width.4pt depth0pt}

\def\IC{\relax\hbox{$\inbar\kern-.3em{\rm C}$}}
\def\IQ{\relax\hbox{$\inbar\kern-.3em{\rm Q}$}}
\def\IR{\relax{\rm I\kern-.18em R}}
 \font\cmss=cmss10 \font\cmsss=cmss10 at 7pt
\def\IZ{\relax\ifmmode\mathchoice
 {\hbox{\cmss Z\kern-.4em Z}}{\hbox{\cmss Z\kern-.4em Z}}
 {\lower.9pt\hbox{\cmsss Z\kern-.4em Z}}
 {\lower1.2pt\hbox{\cmsss Z\kern-.4em Z}}\else{\cmss Z\kern-.4em Z}\fi}

\def\NPB#1#2#3{{\it Nucl.\ Phys.}\/ {\bf B#1} (19#2) #3}
\def\PLB#1#2#3{{\it Phys.\ Lett.}\/ {\bf B#1} (19#2) #3}
\def\PRD#1#2#3{{\it Phys.\ Rev.}\/ {\bf D#1} (19#2) #3}
\def\PRL#1#2#3{{\it Phys.\ Rev.\ Lett.}\/ {\bf #1} (19#2) #3}
\def\PRT#1#2#3{{\it Phys.\ Rep.}\/ {\bf#1} (19#2) #3}
\def\CMP#1#2#3{{\it Commun.\ Math.\ Phys.}\/ {\bf#1} (19#2) #3}
\def\MODA#1#2#3{{\it Mod.\ Phys.\ Lett.}\/ {\bf A#1} (19#2) #3}
\def\IJMP#1#2#3{{\it Int.\ J.\ Mod.\ Phys.}\/ {\bf A#1} (19#2) #3}
\def\NUVC#1#2#3{{\it Nuovo Cimento}\/ {\bf #1A} (#2) #3}
\def\etal{{\it et al.\/}}

\long\def\@caption#1[#2]#3{\par\addcontentsline{\csname
  ext@#1\endcsname}{#1}{\protect\numberline{\csname
  the#1\endcsname}{\ignorespaces #2}}\begingroup
    \small
    \@parboxrestore
    \@makecaption{\csname fnum@#1\endcsname}{\ignorespaces #3}\par
  \endgroup}
\catcode`@=12

\input epsf


\section{Introduction}
\setcounter{footnote}{0}

One of the pressing theoretical issues that confronts the 
Standard Model of particle physics is to understand the 
origins of CP violation.   While the weak interaction is
well-understood to lead to CP violation through complex
phases in the Cabbibo-Kobayashi-Maskawa (CKM) fermion mass matrix,
CP violation may also independently arise from the physics
of the strong interaction.
Ordinarily, one might have assumed the strong interaction
to conserve CP. 
However, the $U(1)$ problem~\cite{Weinberg}
(namely, the inability to interpret the $\eta$ particle as the 
Nambu-Goldstone boson of a spontaneously broken $U(1)$ flavor symmetry)
turns out to require a non-trivial QCD vacuum structure~\cite{tHooft} which in turn
naturally leads to CP violation.
Specifically, one finds that the effective Lagrangian 
describing the strong interaction should be augmented by
an additional CP-violating contribution
\beq
         {\cal L}^{\rm eff}~=~ {\cal L}_{\rm QCD} ~+~ 
            \bar\Theta \, {g^2\over 32\pi^2} F^{\mu \nu}_a \tilde F_{\mu\nu a}~
\eeq
where the $\bar\Theta$ parameter receives two contributions:
\beq
          \bar \Theta ~\equiv ~ \Theta ~+~ {\rm Argdet}\,M~.
\eeq
Here $\Theta$ is the strong-interaction $\Theta$-angle 
reflecting the non-trivial nature of the QCD vacuum,
and ${\rm Argdet}\,M$  (with $M$ denoting the CKM matrix)
is the contribution arising from weak interactions.
However, measurements of
the neutron electric dipole moment place the stringent experimental
bound
\beq
            \bar\Theta ~\lsim~ 10^{-9}~.
\eeq
Explaining the small size of $\bar\Theta$ is the strong CP problem~\cite{Pecceireview}.

The most elegant explanation of the strong CP problem is the
Peccei-Quinn (PQ) mechanism~\cite{PQ}, in which $\bar\Theta$ is set to zero
dynamically as the result of a global, spontaneously broken $U(1)$ Peccei-Quinn
symmetry.  However, associated with this symmetry is a new Nambu-Goldstone boson,
the axion~\cite{Weinbergaxion,Wilczekaxion}, which essentially replaces
the $\bar\Theta$ parameter in the effective Lagrangian.  
This then results in an
effective Lagrangian of the form~\cite{Pecceireview}
\beq
         {\cal L}^{\rm eff}~=~ {\cal L}_{\rm QCD} ~+~ 
         \half \partial_\mu a \partial^\mu a ~+~
             {a\over f_{\rm PQ}}\, \xi\, 
            {g^2\over 32\pi^2} F^{\mu \nu}_a \tilde F_{\mu\nu a}~
\eeq
where $f_{\rm PQ}$ is the axion decay constant, associated with the scale
of PQ symmetry breaking.  Here $\xi$ is a model-dependent parameter 
describing
the PQ transformation properties of the ordinary fermions, and we have not
exhibited further Lagrangian terms which describe axion/fermion couplings.
The mass of the axion is then expected to be of the order
\beq
           m_a ~\sim~ {\Lambda_{\rm QCD}^2\over f_{\rm PQ}}~
\eeq
where $\Lambda_{\rm QCD}\approx 250$ MeV;  likewise, the couplings of axions
to fermions are suppressed by a factor of $1/f_{\rm PQ}$.
Thus, heavier scales for PQ symmetry breaking generally imply lighter axions
which couple more weakly to ordinary matter.

Ordinarily, one might have liked to associate the scale $f_{\rm PQ}$ with
the scale of electroweak symmetry breaking, implying an axion mass
near the electron mass $m_a\approx {\cal O}(10^2)$ keV.
However, experimental searches for such axions have so far 
been unsuccessful~\cite{Kimreview},
and indeed only a narrow allowed window
currently exists
\beq
           10^{8} \, {\rm GeV} ~\lsim ~ f_{\rm PQ} ~\lsim~
           10^{12} \, {\rm GeV} ~.
\label{PQscale}
\eeq
This then implies an axion which is exceedingly light
\beq
           10^{-5} \, {\rm eV} ~\lsim ~ m_a ~\lsim~
           10^{-1} \, {\rm eV} ~,
\label{axionmass}
\eeq
and whose couplings to ordinary matter are exceedingly suppressed.
These bounds generally result from various combinations
of laboratory, astrophysical, and cosmological 
constraints.  In all cases, however, the crucial ingredient is the
correlation between the {\it mass}\/ of the axion and the strength of its
 {\it couplings}\/ to matter, since both are essentially determined by
the single parameter $f_{\rm PQ}$.

In this paper, we point out that this situation may be drastically altered
in theories with large extra spacetime dimensions.
Since their original proposal~\cite{Antoniadis},
such theories have recently received considerable attention
because of their prospects for lowering the fundamental GUT scale~\cite{DDG},
the fundamental Planck scale~\cite{ADD}, and
the fundamental string scale~\cite{WittLykk}.
As a result of these developments, it is now understood that all three
of these scales may be adjusted to arbitrary values, perhaps even
values in the TeV-range, without violating experimental constraints.
Of course, if one lowers these fundamental scales below the 
PQ symmetry-breaking scale in (\ref{PQscale}), then it might seem 
difficult to preserve the axion solution to the strong CP problem. 
In fact, this observation has been used~\cite{burgess}
to argue that the fundamental scales of physics should be taken
at some intermediate scale near the PQ scale.
However, this argument 
assumes that the PQ mechanism itself remains untouched by the presence
of the extra large dimensions.

In this paper, we shall generalize the PQ mechanism
to higher dimensions.  More specifically, we shall consider the
consequences of placing the PQ axion in the ``bulk'' (\ie, perpendicular
to the $p$-brane that contains the Standard Model) so that 
the axion accrues an infinite tower of Kaluza-Klein excitations.  
Placing the QCD axion in the bulk has also been discussed previously
in Refs.~\cite{ADD,CTY}, and is completely analogous to recent ideas
concerning the placement of the right-handed {\it neutrino}\/ in the
bulk~\cite{DDGneutrinos,ADDneutrinos} in order to lower the neutrino seesaw scale   
and obtain neutrino oscillations with light (or even vanishing~\cite{DDGneutrinos}) 
neutrino masses.  

As we shall see, placing the axion in the
bulk has a number of surprising consequences for axion
physics, all of which provide new ways of rendering the axion ``invisible''.
\begin{itemize}
\item  First, in Sect.~2, we shall show that the mass of the axion can be {\it decoupled}\/ 
     from the PQ symmetry-breaking scale, and under certain circumstances
     is essentially set  by the radius $R$ of the bulk:
    \beq
               m_a ~\lsim~ {\cal O} (R^{-1})~.
    \eeq
    Thus, remarkably, the allowed range (\ref{axionmass}) for 
    the axion mass is consistent with radii in the millimeter 
    or sub-millimeter range!
    More importantly, however, this result implies that 
    it is possible to adjust the PQ symmetry breaking scale
    independently of the axion mass
    in order to control the strength of the {\it couplings}\/ of this axion
    to ordinary matter.  This might be useful in order to provide an 
    alternative explanation for an ``invisible'' axion.
  
\item  Second, in Sect.~3, we shall show 
      that the presence of an infinite tower of Kaluza-Klein 
     axion modes can induce the novel phenomenon of 
    {\it laboratory axion oscillations}\/.  These oscillations
    are completely analogous to laboratory neutrino oscillations,
    but we shall find that under certain circumstances
    they lead to a complete and rapid decoherence of the axion field.
   This implies that an axion, once produced in the laboratory,
   will ``decohere'' extremely rapidly.  
    This is therefore a second higher-dimensional phenomenon that can
     contribute to the ``invisibility'' of the axion under certain circumstances.
 
\item  Third, in Sect.~3, we shall also discuss the role played by 
    the excited axion Kaluza-Klein states in axion-mediated processes and
     decays, and propose several experimental methods of 
     detecting their existence.
    These would therefore provide direct experimental tests of 
     the higher-dimensional nature of the axion field.

\item  Finally, in Sect.~4, we shall show that under certain circumstances,
     the presence of the infinite tower of Kaluza-Klein axion modes can 
     actually {\it accelerate}\/ the dissipation of the energy associated with
     cosmological relic axion oscillations.  This implies a weakening
    of the usual four-dimensional
   cosmological relic oscillation bounds on the PQ symmetry-breaking
      scale, which in turn may permit axion/matter couplings to be suppressed
     even more strongly than in the usual four-dimensional case.
      This is therefore a third higher-dimensional feature leading
     to an ``invisible'' axion.  
      
\end{itemize}

Together, these results suggest that ``invisible'' axions can emerge
quite naturally within the context of higher-dimensional theories
with large-radius compactifications, and have significantly different
phenomenologies than they do in four dimensions.
Moreover, many of our results apply to bulk fields in general, and transcend
the specific case of the axion.  This is illustrated in Sect.~6,
where we consider the consequences of extra dimensions for another hypothetical
particle, the so-called Standard-Model dilaton.

\section{A higher-dimensional Peccei-Quinn mechanism}
\setcounter{footnote}{0}

In order to generalize the Peccei-Quinn (PQ) mechanism to higher dimensions,
we will assume that there exists a complex scalar field $\phi$ in
higher dimensions which transforms under a global $U(1)_{\rm PQ}$ symmetry:
\beq
    \phi ~\rightarrow~ e^{i\Lambda} \phi~.
\eeq
This symmetry is assumed to be spontaneously broken by the bulk dynamics
so that
$\langle \phi \rangle = f_{\rm PQ}/\sqrt{2}$, where $f_{\rm PQ}$ is
the energy scale associated with the breaking of the PQ symmetry.
We thus write our complex scalar field $\phi$ in the form
\beq
       \phi ~\approx~ {f_{\rm PQ} \over \sqrt{2}} \,e^{i a/f_{\rm PQ}}
\label{phia}
\eeq
where $a$ is the Nambu-Goldstone boson (axion) field.  
If we concentrate on the case of five dimensions for concreteness,
then the kinetic-energy term for the scalar field takes the form
\beq
  {\cal S}_{\rm K.E.} ~=~ \int d^4x\,dy~ M_s \,\partial_M \phi^\ast
   \partial^M \phi ~=~ 
         \int d^4x\,dy~ M_s \,\half\,
    \partial_M a \partial^M a~
\eeq
where $M_s$ is a fundamental mass scale (\eg, a Type~I string scale),
and where we have neglected the contributions from the radial mode.
Here $x^\mu$ are the coordinates of the four uncompactified spacetime dimensions,
$y$ is the coordinate of the fifth dimension, and the $M$ spacetime
index runs over all five dimensions:  $x^M\equiv(x^\mu,y)$.
Note that there is no mass term for the axion, 
as this would not be invariant under the $U(1)_{\rm PQ}$ transformation
\beq
      a~\rightarrow~ a + f_{\rm PQ}\Lambda~.
\label{PQtrans}
\eeq
Furthermore, as a result of the chiral anomaly, we will
also assume a bulk/boundary coupling of the form 
\beq
     {\cal S}_{\rm coupling}~=~ \int d^4 x \, dy ~{\xi\over f_{\rm PQ}}\,  
   {g^2\over 32\pi^2} \,a \,F^{\mu\nu}_a \tilde F_{\mu\nu a} \,\delta (y)~
\eeq
where $F_{\mu\nu a}$ is the (four-dimensional) QCD field strength describing
the QCD gauge fields which are confined to a four-dimensional subspace
(\eg, a D-brane) located at $y=0$, and where $\xi$ is a model-dependent 
quantity parametrizing the strength
of the axion couplings to matter.  
Thus, our total effective five-dimensional axion 
action takes the form
\beq
   {\cal S}_{\rm eff} ~=~ \int d^4x~dy~ \left[ \,\half \,M_s\,\partial_M a
   \partial^M a ~+~ 
     {\xi\over f_{\rm PQ}}\,  
   {g^2\over 32\pi^2} \,a \,F^{\mu\nu}_a \tilde F_{\mu\nu a} \,\delta (y)\right]~.
\label{lagfive}
\eeq

While we have assumed that the spontaneously broken $U(1)_{\rm PQ}$ is parametrized
by $f_{\rm PQ}$, one still has to address the fact that gravitational effects can 
also break the $U(1)_{\rm PQ}$ symmetry.  In other words, gravitational 
interactions do not respect global symmetries.  Ultimately, this can lead to a 
gravitational contribution to the axion mass which is not necessarily suppressed.  
In this paper, however, we will assume that the gravitational contributions to the axion mass 
are indeed suppressed, and that $U(1)_{\rm PQ}$ remains a valid symmetry even in
the presence of gravitational effects.
For example, such suppression might arise due to 
the suppression of gravitational interactions across a large bulk,
discrete symmetries of the sort that might come from Scherk-Schwarz
compactifications, or other large-radius effects.

In order to obtain an effective four-dimensional theory, our next step is to
compactify the fifth dimension.  For simplicity,
we shall assume that this dimension is compactified on a $\IZ_2$ orbifold of 
radius $R$ where the orbifold action is identified as $y\to -y$.
This implies that the axion
field will have a Kaluza-Klein decomposition of the form
\beq
    a(x^\mu,y)~=~\sum_{n=0}^{\infty} \,a_n(x^\mu) \,\cos\left({n y\over R}\right)  
\label{KKdecomp}
\eeq
where $a_n(x^\mu)\in \IR$ are the Kaluza-Klein modes and 
where we have demanded that the axion field be 
symmetric under the $\IZ_2$ action (in order to have a light zero-mode 
that we can identify with the usual four-dimensional axion).

In principle, we should also allow for the possibility that the axion
field {\it winds}\/ non-trivially around the extra compactified dimension,
with winding number $w$.  This possibility of winding arises because 
$a$ is really only an angular variable, as evident from (\ref{phia}), and
would imply that we should introduce an extra term $w f_{\rm PQ} y /R$ into
the mode expansion (\ref{KKdecomp}).  
Note, in particular, that such windings cannot be removed by global
Peccei-Quinn transformations of the form (\ref{PQtrans}).
However, such a winding term
would not be invariant under the orbifold symmetry $y\to -y$, and thus
only the unwound configuration $w=0$ survives the orbifold projection.
Even if we were to compactify the axion field on a circle rather than an orbifold,
such a term would contribute only an overall additive constant
to the resulting effective potential for the axion field, and would
not change any of the subsequent physics.  
It is therefore sufficient to restrict our attention to
the unwound configuration with $w=0$. 
 
It is also interesting to note that with respect to the four-dimensional
Kaluza-Klein axion modes $a_n$, the Peccei-Quinn transformation (\ref{PQtrans})
takes the form
\beq
            \cases{      a_0 \to a_0 + f_{\rm PQ} \Lambda & \cr  
                         a_k \to a_k~ ~~~~~~~~~~~~~{\rm for~all}~k>0~. & \cr}
\label{PQtransmodes}
\eeq
Thus, we see that only $a_0$ serves as the true axion transforming 
under the PQ transformation, while the excited Kaluza-Klein modes $a_k$ 
remain invariant.

Substituting (\ref{KKdecomp}) into (\ref{lagfive}) and integrating over the
fifth dimension, we then obtain an
effective four-dimensional Lagrangian density
\beq
   {\cal L}_{\rm eff} ~=~ {\cal L}_{\rm QCD} ~+~ \half 
   \sum_{n=0}^{\infty} (\partial_\mu a_n)^2
            ~-~ \half \sum_{n=1}^{\infty} {n^2\over R^2} a_n^2 
                ~+~
        {\xi\over \hat f_{\rm PQ}}\, {g^2\over 32\pi^2}
    \left(\sum_{n=0}^{\infty} r_n a_n\right)~
            F^{\mu\nu}_a \tilde F_{\mu\nu a}~
\label{lag}
\eeq
where  
\beq
        r_n ~\equiv ~\cases{ 1 & if $n=0$\cr
                          \sqrt{2} &  if $n>0$ ~.\cr}
\label{rdefs}
\eeq
Note that in order to obtain (\ref{lag}), we must 
individually rescale each of the Kaluza-Klein modes $a_n$ in 
order to ensure that they have canonically normalized kinetic-energy terms. 
It is this that produces the relative rescaling coefficients $r_n$ in (\ref{rdefs}).
We have also defined $\hat f_{\rm PQ}\equiv (V M_s)^{1/2} f_{\rm PQ}$,
where $V$ is the volume of our compactified space.
For $\delta$ extra dimensions, this definition generalizes to
$\hat f_{\rm PQ}\equiv (V M_s^\delta)^{1/2} f_{\rm PQ}$.
For example, assuming a $\delta$-dimensional toroidal compactification 
implies $V=(2\pi R)^\delta$, resulting
in the relation 
\beq
      \hat f_{\rm PQ} ~\equiv~  (2\pi R M_s)^{\delta/2} \,f_{\rm PQ}~.
\label{fhatdef}
\eeq

Note that while $f_{\rm PQ}$ sets the overall mass scale for
the breaking of the Peccei-Quinn symmetry, it is the volume-renormalized
quantity $\hat f_{\rm PQ}$ that parametrizes the coupling between the axion
and the gluons.
In general, since $M_s\gg R^{-1}$,
we find that $\hat f_{\rm PQ} \gg f_{\rm PQ}$.
Therefore, as pointed out in Ref.~\cite{ADD}, this
volume-renormalization of the brane/bulk coupling can be used
to obtain sufficiently suppressed axion/gauge-field couplings
even if $f_{\rm PQ}$ itself is taken to be relatively small.
In other words, even if we demand that $\hat f_{\rm PQ}$ be in the approximate
range $10^{8} \,{\rm GeV} \lsim \hat f_{\rm PQ}\lsim  10^{12}$ GeV, the fundamental
Peccei-Quinn symmetry-breaking
scale $f_{\rm PQ}$ can be substantially reduced, potentially all the 
way to the TeV-range.  This volume suppression is thus one 
higher-dimensional way~\cite{ADD} of avoiding the need
for a high fundamental Peccei-Quinn scale $f_{\rm PQ}$. 

Two comments are important at this stage.
First, in toroidally compactified higher-dimensional 
theories with reduced Planck scales, the higher-dimensional
Planck scale $M_\ast$ and the usual four-dimensional Planck 
scale $M_{\rm Planck}$
are related to each other via~\cite{ADD}
\beq
          M_{\rm Planck} ~=~ \,(2\pi R M_\ast)^{n/2} \, M_\ast~
\eeq
where $n$ is the total number of extra spacetime dimensions in the bulk.
Identifying $M_\ast\sim M_s$, we therefore see that the $n$-dimensional
volume factor $(2\pi RM_s)^{n/2}$ must already be adjusted in order 
to account for the difference between $M_s\sim {\cal O}$(TeV)
and $M_{\rm Planck}\sim {\cal O}(10^{19}\,{\rm GeV})$.
If we were to take $\delta=n$ for the current axion case, this would
imply either that $\hat f_{\rm PQ}\sim M_{\rm Planck}$ (which would presumably
overclose the universe), or that $M_s\ll {\cal O}$(TeV) (which would clearly 
violate current experimental bounds).  Therefore, if we assume an 
isotropic compactification with all radii taken equal, we see 
that an intermediate scale $\hat f_{\rm PQ}$
can be generated via (\ref{fhatdef}) only if we have $\delta<n$.
In other words, under these assumptions, the
axion must be restricted to a {\it subspace}\/ of the full higher-dimensional bulk.
This has already been pointed out in Ref.~\cite{CTY}, and is analogous to
similar restrictions that arise in the case of higher-dimensional 
neutrinos~\cite{DDGneutrinos,ADDneutrinos}.
    
Our second comment concerns the relevance of the mass scale 
$\hat f_{\rm PQ}$ that is generated by this volume factor.
Of course, it is apparent from (\ref{lag}) 
that $\hat f_{\rm PQ}$ (rather than $f_{\rm PQ}$) sets 
the scale for couplings between gauge fields and
individual axion modes $a_k$.  However, we have also seen 
in (\ref{lag}) that the gauge fields couple not to an individual
axion mode $a_k$, but rather to the linear superposition
\beq
         a' ~\equiv~  {1\over \sqrt{N}} \,
              \sum_{n=0}^{n_{\rm max}} \, r_n a_n ~=~
              {1\over \sqrt{N}}\,
         \left(a_0 + \sqrt{2} \sum_{n=1}^{n_{\rm max}} a_n\right)~
\label{apdef}
\eeq
where 
\beq
             N ~\equiv~ 1+2n_{\rm max}~.
\label{Ndef}
\eeq
Here $n_{\rm max}$ is a cutoff, determined according to the underlying
mass scale $M_s$ (which sets the limit of validity of
our higher-dimensional effective field theory\footnote{
       Note that the heavy modes with
       masses of order $M_s$ can be treated within field theory
       along the lines discussed in Ref.~\cite{BKNY}, and 
       within string theory along the lines discussed in Ref.~\cite{DM}.}).   
Taking $n_{\rm max}\approx R M_s\gg 1$, we then find that
the axion/gluon coupling in (\ref{lag}) takes the form
\beq
           {\sqrt{N} \over \hat f_{\rm PQ}} \, a' \,
          F^{\mu\nu}_a \tilde F_{\mu\nu a} ~\approx~
           {1 \over \sqrt{\pi} f_{\rm PQ}} \, a' \,
          F^{\mu\nu}_a \tilde F_{\mu\nu a}~.
\eeq
Thus it is actually $f_{\rm PQ}$, rather than $\hat f_{\rm PQ}$, that
sets the scale for axion couplings involving the entire Kaluza-Klein
linear superposition $a'$.  In other words, the effects of the volume
factor in (\ref{fhatdef}) are cancelled by the normalization of
the Kaluza-Klein linear superposition $a'$.
Of course, this is expected from the perspective of the higher-dimensional
theory in which $f_{\rm PQ}$ is the only fundamental mass scale, and
indeed this cancellation of the volume factor persists for any number
of extra spacetime dimensions.
Because we expect $f_{\rm PQ}\ll \hat f_{\rm PQ}$ in scenarios
with large extra spacetime dimensions,
we see that axion couplings involving the linear superposition
$a'$ are relatively strong, and pose a serious threat to 
the invisibility of the higher-dimensional axion.  We shall discuss how 
this problem may be overcome in subsequent sections.
 
Similar observations also apply for axion couplings to Standard-Model fermions.
The invariance under the $U(1)_{\rm PQ}$ transformation 
$a\rightarrow a+ f_{\rm PQ}\Lambda$ 
implies that axions can be at most derivatively coupled to
fermions carrying a PQ charge.  If we assume that these fermions
are also restricted to the D-brane at $y=0$ (as would be the case
for all Standard-Model fermions), then this axion/fermion coupling 
is restricted to take the form 
\beq
    {\cal S}_{a\psi\psi}~\sim ~ \int d^4x \, {1\over \hat f_{\rm PQ}} 
          (\partial_\mu a|_{y=0}) ({\bar\psi} \gamma^\mu \gamma^5 \psi)~.
\eeq
Here $a|_{y=0}$ is the full five-dimensional bulk axion field evaluated
at $y=0$, and the bulk/brane coupling strength $\hat f_{\rm PQ}$    
is defined in (\ref{fhatdef}).
Note that $\hat f_{\rm PQ}$ is the same volume-rescaled  
axion decay constant that parametrizes the couplings to the gauge fields,
with the volume factor emerging just as for the axion/gauge couplings.
Using the Kaluza-Klein decomposition (\ref{KKdecomp}),  
the action then becomes
\beq
    {\cal S}_{a\psi\psi}~\sim~ {1\over \hat f_{\rm PQ}} 
      \,\sum_{n=0}^{\infty} r_n \,\int d^4x \, (\partial_\mu a_n )\, ({\bar\psi} 
           \gamma^\mu\gamma^5 \psi)~\sim~
    {1\over f_{\rm PQ}} 
      \int d^4x \, (\partial_\mu a' )\, ({\bar\psi} \gamma^\mu\gamma^5 \psi)~,
\label{axfermion}
\eeq
and we see that once again the entire Kaluza-Klein
linear superposition $a'$ couples to the charged fermions. 
This is completely analogous to the situation in (\ref{lag}) for the gauge fields.
It is important to stress that this need not have been the result 
from a purely four-dimensional perspective.
Indeed, given the Peccei-Quinn transformation properties (\ref{PQtransmodes}),
we see that it is only the zero-mode $a_0$ which requires a derivative
coupling to fermions;  the other axion modes $a_k$ are {\it a priori}\/ free
to have non-derivative couplings.  It is therefore only the 
higher-dimensional structure of the axion field that forces
all of the axion modes to have identical derivative couplings to
charged fermions.
Moreover, we see from (\ref{axfermion}) that while the mass scale
for the couplings of individual Kaluza-Klein axion 
fermions to fermions is set by $\hat f_{\rm PQ}$, the mass scale 
for the coupling of the full linear superposition 
$a'$ to fermions is set by $f_{\rm PQ}$.

Let us now proceed to verify that this 
higher-dimensional Peccei-Quinn mechanism still cancels the CP-violating phase,
and use this to calculate the {\it mass}\/ of the axion.
Given the Lagrangian (\ref{lag}), we observe that $a'\sim \sum_{n=0}^\infty r_n a_n$ 
serves as the overall quantity that parametrizes the size of CP symmetry breaking.
Applying the one-instanton dilute-gas approximation, it is straightforward
to show that 
\beq
      \langle F^{\mu\nu}_a \tilde F_{\mu\nu a}\rangle ~=~
      - \Lambda_{\rm QCD}^4\, \sin\left(
           {\xi\over \hat f_{\rm PQ}} 
        \sum_{n=0}^{\infty} r_n a_n + \bar \Theta \right)~.
\eeq
This gives rise to an effective potential for the axion modes in the QCD vacuum
\beq
    V(a_n)  ~=~ \half \sum_{n=1}^{\infty} {n^2\over R^2} a_n^2 
        ~+~ {g^2\over 32\pi^2}\, \Lambda_{\rm QCD}^4
       \left[ 1 - \cos\left( 
           {\xi\over \hat f_{\rm PQ}} 
        \sum_{n=0}^{\infty} r_n a_n + \bar \Theta \right)\right]~.
\eeq
In order to exhibit the Peccei-Quinn mechanism,
we now minimize the axion effective potential,
\beq
      {\partial V\over \partial a_n} ~=~
           {n^2\over R^2} a_n ~+~  r_n
              {\xi\over \hat f_{\rm PQ}}\, {g^2\over 32\pi^2}\,
            \Lambda_{\rm QCD}^4\, \sin\left( {\xi\over \hat f_{\rm PQ}} 
           \sum_{n=0}^{\infty}r_n a_n
             + \bar\Theta \right)~=~0~,
\eeq
yielding the unique solution
\beqn
     \langle a_0 \rangle &=&  
         {\hat f_{\rm PQ}\over\xi}(-\bar\Theta + \ell\pi) ~,~~~ \ell\in 2\IZ~\nonumber\\
         \langle a_k \rangle &=&  0~~~~       {\rm for~all} ~k>0~.
\label{PQsoln}
\eeqn
Note that while any value $\ell\in \IZ$ provides an extremum of the potential,
only the values $\ell\in 2\IZ$ provide the desired  
 {\it minimum}\/ of the potential.
Thus, this higher-dimensional Peccei-Quinn mechanism continues to
solve the strong CP problem:  
we see that $a_0$ is the usual Peccei-Quinn axion which solves the strong CP problem 
by itself by cancelling the $\bar\Theta$ angle,
while all of the excited Kaluza-Klein axions $a_k$ for $k>0$ have 
vanishing VEVs.
This makes sense, since only $a_0$ is a true massless Nambu-Goldstone field from
the four-dimensional perspective of (\ref{lag}).
This is also evident from the Peccei-Quinn transformation properties
(\ref{PQtransmodes}). 

However, as we shall now show, these excited Kaluza-Klein 
axion states nevertheless have a drastic effect on the axion mass matrix.
In order to derive the mass matrix, we consider
the local curvature of the effective axion potential around its minimum:
\beq
     {\cal M}^2_{nn'} ~\equiv~  
        {\partial^2 V\over \partial a_n \partial a_{n'}} 
          \Biggl|_{\langle a \rangle}~.
\label{massderivatives}
\eeq 
 From this we obtain
\beq
     {\cal M}^2_{nn'} ~\equiv~ {n^2\over R^2} \delta_{nn'} ~+~ 
      \xi^2 {g^2\over 32\pi^2} \, 
     {\Lambda_{\rm QCD}^4 \over \hat f_{\rm PQ}^2} \,r_n r_{n'}\,\cos\left(
           {\xi\over \hat f_{\rm PQ}}
       \sum_{n=0}^\infty r_n a_n
          + \bar\Theta\right)\Biggl |_{\langle a \rangle}~,
\eeq
and in the vicinity of the minimum (\ref{PQsoln}) this becomes
\beq
          {\cal M}^2_{nn'} ~=~ {n^2\over R^2} \delta_{nn'} ~+~ 
           \xi^2 {g^2\over 32\pi^2} {\Lambda_{\rm QCD}^4\over \hat f_{\rm PQ}^2} 
              \,r_n r_{n'}~.
\label{matrixalg}
\eeq
Let us now define
\beqn
     m^2_{\rm PQ} &\equiv& \xi^2 {g^2\over 32\pi^2} {\Lambda_{\rm QCD}^4 
          \over \hat f^2_{\rm PQ}}\nonumber \\
     y &\equiv& {1\over m_{\rm PQ} R}~.
\label{defs}
\eeqn
Thus $m_{\rm PQ}$ is the expected mass 
that the axion would ordinarily have taken in four dimensions
(depending on $\hat f_{\rm PQ}$ rather than $f_{\rm PQ}$ itself),
and $y$ is the ratio of the scale of the extra dimension to $m_{\rm PQ}$.
Our mass matrix then takes the form
\beq
          {\cal M}^2 ~=~ 
            m_{\rm PQ}^2 \, \left( r_n r_{n'} ~+~ y^2 \,n^2 \,\delta_{nn'} \right)~,
\eeq
or equivalently
\beq
      \calM^2 ~=~  m_{\rm PQ}^2 \, \pmatrix{   1 & \sqrt{2} & \sqrt{2} & \sqrt{2} &\ldots \cr
                  \sqrt{2} & 2+y^2 & 2 & 2& \ldots \cr
                  \sqrt{2} & 2 & 2+4 y^2 & 2& \ldots \cr
                  \sqrt{2} & 2 & 2 & 2+9y^2& \ldots \cr
                  \vdots & \vdots & \vdots & \vdots  & \ddots\cr}~.
\label{matrix}
\eeq
Note that the usual Peccei-Quinn case corresponds to the upper-left 
$1\times 1$ matrix, leading to the expected result $\calM^2= m^2_{\rm PQ}$.
Thus, the additional rows and columns reflect the extra Kaluza-Klein states, 
and their physical effect is to pull the lowest eigenvalue of this matrix
away from $m^2_{\rm PQ}$.

Deriving the condition for the eigenvalues of this matrix is straightforward.
Let us denote the eigenvalues of this matrix as $\lambda^2$
rather than $\lambda$ because this is a (mass)$^2$ matrix.
We then find that the eigenvalues are given as the solutions to the  
transcendental equation
\beq
         {\pi \tilde \lambda\over y} \, 
            \cot\left( {\pi \tilde \lambda\over y}\right) ~=~
            \tilde \lambda^2~
\label{constraint}
\eeq
where we have defined the dimensionless eigenvalue
\beq
                  \tilde \lambda ~\equiv~ \lambda / m_{\rm PQ} ~.
\label{dimeig}
\eeq
In terms of dimensionful quantities, this transcendental equation
takes the equivalent form\footnote{
    Interestingly, this eigenvalue equation is identical to that
     which emerges~\cite{DDGneutrinos} 
     when the right-handed neutrino $\nu_R$ is placed in the bulk,
      with the mass scale $m_{\rm PQ}$ in the axion case 
     corresponding to the Dirac coupling $m$ in
     the neutrino case.
     This implies that there is a formal
    relation between the Kaluza-Klein axion modes and the Kaluza-Klein
    neutrino modes.
     Remarkably, this correspondence exists
    even though the axion and right-handed neutrino have different spins, 
     and even though the mechanisms for mass generation are completely
    different in the two cases.  
     Moreover, in Sect.~6,
    we shall demonstrate that the same mass matrix and eigenvalue
    equation also emerge when the Standard-Model {\it dilaton}
    is placed in the bulk of extra spacetime dimensions.
    This suggests that many of the higher-dimensional
    phenomena to be discussed in this paper (such as 
     laboratory and cosmological relic axion oscillations)
    may have a correspondingly general phenomenology that is equally
    applicable to neutrinos, dilatons, as well as other bulk fields that
    transform as singlets under the Standard-Model gauge group.}  
\beq
      \pi R \lambda \,\cot(\pi R\lambda) ~=~  {\lambda^2\over m_{\rm PQ}^2 } ~.
\label{finalconstraint}
\eeq
For each eigenvalue $\lambda$, the corresponding
normalized mass eigenstate $\hat a_\lambda$ is exactly given by
\beq
         \hat a_\lambda ~=~ \sum_{k=0}^\infty
               \,U_{\lambda k} \,a_k
\label{masseig}
\eeq
where $a_k$ are the Kaluza-Klein axion modes given in (\ref{KKdecomp})
and
where $U_{\lambda k}$ is the unitary matrix that diagonalizes ${\cal M}^2$.
This matrix is given by
\beq
          U_{\lambda k} ~\equiv~  
            \left({r_k \,\tilde \lambda^2 \over 
           \tilde \lambda^2 - k^2 y^2 }\right) \, A_\lambda
\label{Umatrix}
\eeq
where 
\beq
           A_\lambda ~\equiv~ {\sqrt{2}\over \tilde \lambda}\, 
           \left(\tilde \lambda^2  +  1 + \pi^2/y^2  \right)^{-1/2}~.
\label{Adef}
\eeq
Note that the unitarity of the matrix $U$ implies 
that $\sum_\lambda |U_{\lambda 0}|^2=1$,
which in turn implies 
\beq
          \sum_\lambda  \, A_\lambda^2~=~1~.  
\eeq
For future reference, we also record another useful identity which 
will be proven in Sect.~4:
\beq
          \sum_\lambda  \, \tilde \lambda^2 \, A_\lambda^2 ~=~ 1~.
\label{identity}
\eeq
Finally, combining (\ref{constraint})
and (\ref{Umatrix}),
it is straightforward to show that
\beq
           \sum_{k=0}^\infty  \, r_k \, U_{\lambda k} ~=~ 
                  \tilde \lambda^2 \,A_\lambda~.
\label{otheridentity}
\eeq
Note that all three of these identities hold for all values of 
$y\equiv (m_{\rm PQ} R)^{-1}$.
This in turn allows us to rewrite (\ref{apdef}) in the form
\beq
       a'~=~ {1\over \sqrt{N}}  \,\sum_\lambda \,
            \tilde \lambda^2 \,A_\lambda \,\hat a_\lambda~.
\label{apdeflambda}
\eeq

We can check that (\ref{finalconstraint}) makes sense in the limit $R\to 0$.
In this limit, the Kaluza-Klein states become infinitely heavy and decouple;
thus we should be left with the lightest eigenvalue $\lambda= m_{\rm PQ}$.
And this is indeed what happens:  as $R\to 0$, we see
           $\pi R \lambda \cot(\pi R\lambda) \to 1$,
whereupon we obtain $\lambda= m_{\rm PQ}$, with all other eigenvalues infinitely
heavy.
As $R$ becomes larger, the effect of the extra large dimension is felt
through a reduction of this lowest eigenvalue.  
Thus, as $R$ increases, the mass of the lightest axion decreases.

\begin{figure}[ht]
\centerline{
      \epsfxsize 3.9 truein \epsfbox {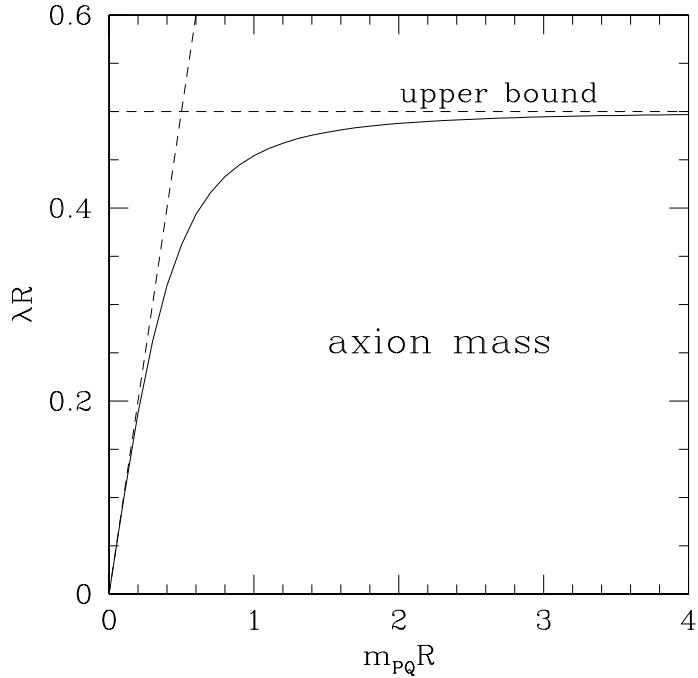}}
\caption{ The mass of the axion zero-mode as a function
        of the dimensionless product $m_{\rm PQ}R\equiv y^{-1}$, where
         $m_{\rm PQ}\sim \Lambda_{\rm QCD}^2/\hat f_{\rm PQ}$.
         Although this reproduces the
         expected result $m_a\approx m_{\rm PQ}$ when $m_{\rm PQ}\ll R^{-1}$
         (corresponding to the diagonal dashed line $\lambda R=m_{\rm PQ}R$),
         we see that the axion mass is strictly bounded by
         the inverse radius in higher dimensions,
        with the precise value of $m_{\rm PQ}$ essentially
        decoupling for $m_{\rm PQ} \gsim \half R^{-1}$.
        This implies that in higher dimensions,
       the Peccei-Quinn scale $f_{\rm PQ}$
        can be adjusted arbitrarily within this parameter range
         without affecting the axion mass.} 
\label{massfig}
\end{figure}

One important consequence of (\ref{finalconstraint}) is that the
lightest axion mass eigenvalue $m_a$ is strictly bounded by the radius
\beq
          m_a ~\leq~ \half \,R^{-1}~.
\label{majresult1}
\eeq
This result (\ref{majresult1}) holds {\it regardless}\/ of the value of $m_{\rm PQ}$.
Thus, 
in higher dimensions,
we see that 
when $m_{\rm PQ} \gsim \half R^{-1}$, 
the size of the axion mass is set by the
radius $R$ and not by the Peccei-Quinn scale $f_{\rm PQ}$. 
In Fig.~\ref{massfig} we show the value of the axion mass $m_a$ as a function
of $y^{-1}\equiv m_{\rm PQ} R$.
Of course, for $m_{\rm PQ} R\to 0$ (corresponding to either $R\to 0$ or 
$m_{\rm PQ}\to 0$), we see that we indeed have the expected result
$m_a\approx m_{\rm PQ}$.
This is indicated by the diagonal dashed line in Fig.~\ref{massfig}.
However, as $m_{\rm PQ} R$ increases, we see that the axion mass departs
from this expected linear behavior, and instead is bounded by the
inverse radius of the extra spacetime dimension. 
In fact, from Fig.~\ref{massfig}, we see that we can approximate the
mass of the axion as
\beq
          m_a ~\approx~ {\rm min}\,(\half R^{-1}, m_{\rm PQ})~.
\eeq
Thus the mass of the axion is determined solely by the radius of the extra
spacetime dimension when $\half R^{-1}\lsim m_{\rm PQ}$.

As a result of this unexpected higher-dimensional behavior for the axion
mass, we see that 
when $m_{\rm PQ} \gsim \half R^{-1}$,
the Peccei-Quinn scale $f_{\rm PQ}$ 
essentially {\it decouples}\/ from the axion mass!
Indeed, as long as $m_{\rm PQ}\gsim \half R^{-1}$,
we see that $m_a\leq \half R^{-1}$
 {\it regardless}\/ of the specific sizes of $m_{\rm PQ}$ or 
$\Lambda_{\rm QCD}$. 

This observation has a number of interesting implications.
First, given (\ref{majresult1}), we see that an axion mass 
in the allowed range (\ref{axionmass}) is already achieved for 
$R$ in the submillimeter range, independently of $m_{\rm PQ}$!
This therefore provides further motivation for such submillimeter
extra dimensions. 

Second, and even more importantly, we 
now have the surprising result that $m_{\rm PQ}$ can 
still be lowered or raised arbitrarily without upsetting the 
constraint (\ref{axionmass}), provided $m_{\rm PQ}\gsim \half R^{-1}$. 
In other words, having already satisfied the axion {\it mass}\/
constraints by appropriately choosing the value of $R$,  
we are now essentially free to tune $m_{\rm PQ}$ 
(or equivalently the fundamental Peccei-Quinn symmetry breaking
scale $f_{\rm PQ}$)  in such a way as to 
weaken the axion couplings to matter 
to whatever values are required to make the axion sufficiently
invisible. 
This may therefore provide a new method of obtaining an invisible 
axion.

\section{Laboratory axion oscillations}
\setcounter{footnote}{0}

In this section we discuss the novel 
possibility of {\it laboratory axion oscillations}\/.
Note that these are not the traditional cosmological {\it relic}\/ 
axion oscillations (which will be discussed in Sect.~4),
but rather ``laboratory'' axion oscillations that now arise in
higher dimensions because the
physical Peccei-Quinn axion $a_0$ is no longer a mass eigenstate.
These laboratory axion oscillations are therefore completely analogous
to laboratory neutrino oscillations, which similarly arise because the
neutrino gauge eigenstates differ from the neutrino mass eigenstates. 
In the present axion case,
such oscillations are possible because 
of the non-diagonal nature of the axion mass matrix (\ref{matrix}).
Therefore, as it propagates, it is possible for the 
four-dimensional axion zero-mode $a_0$
to oscillate into any of the higher-frequency axion modes
in its Kaluza-Klein tower.
Indeed, as we shall see, these oscillations can provide yet another
mechanism that may contribute to the ``invisibility'' of the axion.

\subsection{Laboratory oscillations of the axion zero-mode $a_0$}

It is straightforward to calculate these oscillation probabilities
in terms of the mass mixing matrix (\ref{matrix}) 
and the $U$-matrix (\ref{Umatrix}) that diagonalizes it.
 From a four-dimensional perspective,
we see from (\ref{PQtransmodes}) that only the zero-mode $a_0$ 
serves as a {\it bona-fide}\/ axion transforming under the PQ transformation.
Therefore, let us first calculate
the probability that this four-dimensional axion
$a_0$ oscillates into any of its corresponding Kaluza-Klein
excitations as it propagates, or conversely the probability 
that the four-dimensional axion $a_0$ is preserved as a function of time.
Assuming that the axion 
is given an initial highly relativistic momentum
$p$, we find that the probability that $a_0$ is preserved is given by
\beqn
            P_{0\to 0}(t) &=&  \left|
           \sum_\lambda \, A_\lambda^2  \,e^{-\Gamma_\lambda t/2}\,
     \exp\left( {i\lambda^2 t\over 2 p}\right) \right|^2~\nonumber\\
      &=& \sum_\lambda A_\lambda^4  e^{-\Gamma_\lambda t}
         ~+~ 2\sum_{\lambda' < \lambda}
            \,A_\lambda^2  A_{\lambda'}^2  \,e^{-(\Gamma_\lambda + \Gamma_{\lambda'})t/2}\,
          \cos\left( {\lbrack \lambda^2 - (\lambda')^2\rbrack t \over 2 p}\right)~.~~~~~
\label{p001}
\eeqn    
Here $A_\lambda$ is defined in (\ref{Adef}), and $\Gamma_\lambda$ is
the decay width of the corresponding Kaluza-Klein axion. 
Thus, even though $P_{0\to 0}=1$ at the initial time $t=0$, 
we see that this probability decreases at later times and 
ultimately oscillates around a period-averaged value
\beq
         \langle P_{0\to 0}(t) \rangle ~=~ 
         \sum_\lambda \, A_\lambda^4\,e^{-\Gamma_\lambda t}~
\label{p00}
\eeq
which itself diminishes exponentially with time.
Note that the calculations leading to these results are similar
to the higher-dimensional
neutrino oscillation calculations in Ref.~\cite{DDGneutrinos}.

It is important at this stage to separate two effects which influence
the axion preservation probability.  The first is the oscillation
itself, which arises due to the non-trivial axion mass matrix and which 
reflects the mixing of the excited Kaluza-Klein axion states.
It is this oscillation which is our focus in this section.
By contrast, the second effect is axion {\it decay}\, as reflected
in the decay widths $\Gamma_\lambda$.  
In general, these decay widths result 
from the dominant decay mode $\hat a_\lambda\to \gamma\gamma$,
and therefore 
scale as $\Gamma_\lambda\approx \lambda^3/\hat f_{\rm PQ}^2$. 
This implies that
for the lowest Kaluza-Klein eigenvalues and sufficiently small times,
the product $\Gamma_\lambda t$ is typically extremely small. 
This is in accordance with 
our expectation that the usual four-dimensional axion is extremely stable.
For example, taking
$\lambda_0\approx m_{\rm PQ}\approx 10^{-5}$~eV,
we see that $\Gamma_0 t\ll 1$ for all times $t\lsim 10^{48}$ seconds.
This upper limit exceeds the age of the universe by 20 orders of magnitude.
Therefore, particularly for the lowest Kaluza-Klein eigenvalues, it is safe
to neglect these decay widths entirely, and concentrate solely
on the oscillations. 
For example, since $A_\lambda^4$ decreases rapidly as a function
of $\lambda$, 
only the lowest eigenvalues dominate the sum in (\ref{p00}).
We therefore find that we can approximate
\beq
         \langle P_{0\to 0}(t) \rangle ~\approx ~ 
        \sum_\lambda \, A_\lambda^4~,
\label{p00new}
\eeq
which is independent of time.
We stress, however, that the decay widths $\Gamma_\lambda$
grow rapidly as a function of the mass of the Kaluza-Klein
eigenstate $\hat a_\lambda$.  
We have therefore included these decay widths in (\ref{p001}) and (\ref{p00})
for completeness, and will discuss the effects that they induce 
more carefully  at the end of this section.

\begin{figure}[th]
\centerline{
      \epsfxsize 3.1 truein \epsfbox {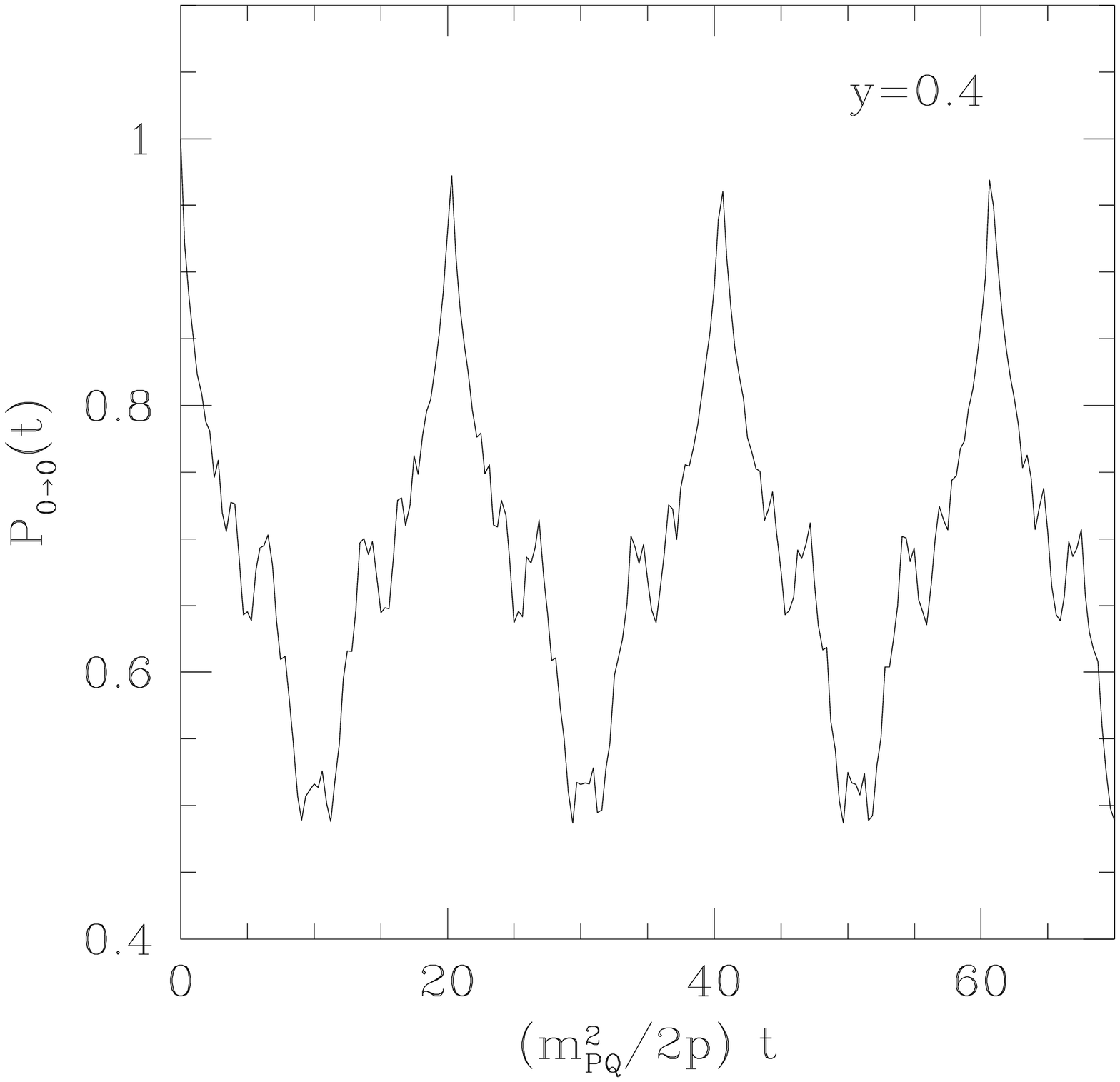}
      \hskip 0.2 truein
      \epsfxsize 3.1 truein \epsfbox {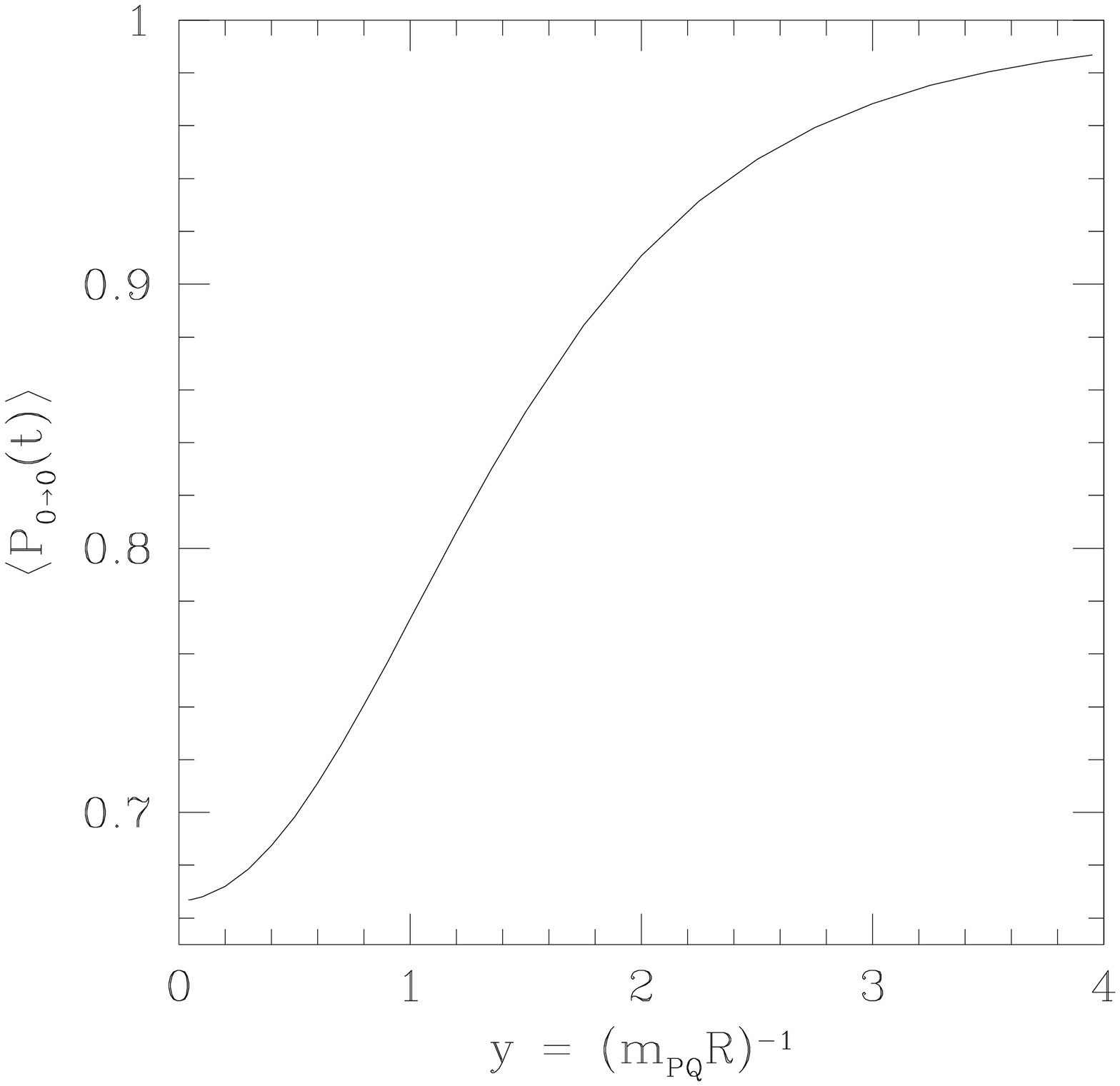}}
\caption{
     Higher-dimensional axion oscillations, as discussed
     in the text.  (a)  The probability $P_{0\to 0}(t)$ 
     that the axion zero-mode is preserved as a function of time, 
     taking a reference value
     $y\equiv (m_{\rm PQ}R)^{-1}=0.4$.  (b)  The period-averaged
     preservation probability $\langle P_{0\to 0}(t)\rangle$
     as a function of $y$.  The limit $y\gg 1$ corresponds to
     the usual four-dimensional case, while the opposite $y\ll 1$
      limit corresponds to an extremely large extra dimension with
      very light Kaluza-Klein states.} 
\label{oscillplots}
\end{figure}

In Fig.~\ref{oscillplots}(a), we have plotted the behavior of $P_{0\to 0}(t)$ 
as a function of time, taking a reference value $y\equiv (m_{\rm PQ} R)^{-1}=0.4$.
Note that the jaggedness of the probability curve reflects the 
multi-component nature of the oscillation in which many different 
individual Kaluza-Klein oscillations interfere with incommensurate
phases.
Although this oscillation clearly leads to both axion deficits and axion
regenerations, we see that while the axion regenerations are nearly total,
the axion deficits are not total.
This is in marked constrast to the results from a simple
two-state oscillation. 
We also observe that these oscillations 
are approximately periodic, 
with a wavelength set by the lowest-lying eigenvalue difference. 
This is because it is the lowest-lying Kaluza-Klein axions that 
play the dominant role in producing this oscillation.
As a result of this fact, we see that 
$\langle P_{0\to 0}(t)\rangle$ is effectively constant as
a function of time, in accordance with (\ref{p00new}).
Indeed, the interpretation of this oscillation is completely analogous to that
given in Ref.~\cite{DDGneutrinos} for higher-dimensional neutrino oscillations.
We shall discuss the possibilities for experimentally
detecting such oscillations at the end of this section.

In Fig.~\ref{oscillplots}(b), we have plotted the time-averaged probability
$\langle P_{0\to 0}(t)\rangle$
given in (\ref{p00new})
as a function of $y\equiv (m_{\rm PQ} R)^{-1}$.
For $y\gg 1$ (corresponding to the usual four-dimensional limit),
we see that
$\langle P_{0\to 0}(t)\rangle \to 1$, as expected, reflecting the fact that our 
single axion field cannot oscillate because its Kaluza-Klein states are
infinitely heavy and essentially decouple.
More interestingly, however, we see that in the $y\ll 1$ limit (corresponding
to extremely large radii or equivalently a quasi-continuous spectrum of light
Kaluza-Klein modes), the oscillation probability (\ref{p00new}) approaches 
a fixed value
\beq
        \lim_{y\to 0} \,\langle P_{0\to 0}(t)\rangle  ~=~ 
        \lim_{y\to 0}  \,\sum_\lambda \, A_\lambda^4 ~=~ {2\over 3}~.
\eeq
Thus, for extremely large radii, we expect to see on average only 2/3 of
the axion flux that would have appeared in the four-dimensional case.

\subsection{Laboratory oscillations of the axion superposition $a'$}

For many practical purposes, there exists a different probability 
that may be more relevant as a measure of laboratory axion oscillations.
As we have seen in Sect.~2, Standard-Model gauge bosons 
and fermions generically couple not to $a_0$, but rather to 
the linear superposition $a'$ given in (\ref{apdef}).  
Thus, in any laboratory process that produces axions or is mediated by axions,
a more crucial oscillation
probability is the probability $P_{a'\to a'}(t)$ that this particular
linear combination $a'$ is preserved as a function of time.
Indeed, we have already seen in Sect.~2 that while the Standard-Model
couplings to individual axion modes scale as $1/\hat f_{\rm PQ}$
and hence are already somewhat ``invisible'',
the couplings to $a'$ scale 
as $1/f_{\rm PQ}$ and hence are significantly larger.
Such couplings therefore pose the largest immediate threat 
to axion invisibility.

The calculation of the probability $P_{a'\to a'}(t)$
proceeds in an analogous manner.
In general, the amplitude for an axion transition $a_k\to a_\ell$ is given by
\beq
        A_{k\to \ell}(t) ~=~ \sum_\lambda \,U_{\lambda \ell} 
            \,U^\ast_{\lambda k} \,e^{-\Gamma_\lambda t/2}\,e^{-i \lambda^2 t/2p}~
\eeq
where $U_{\lambda k}$ are the (real) unitary matrix elements defined in (\ref{Umatrix})
and where we have omitted an overall $(k,\ell)$-independent phase.
The probability that $a'$ is preserved as a function of time is
therefore given by
\beq
     P_{a'\to a'}(t) ~=~ {1\over N^2} \, \left|  \sum_{k,\ell =0}^\infty   
              r_k r_\ell  \, A_{k\to \ell}(t) \right|^2~
\label{newprob}
\eeq
where the normalization factor $N$ is  given in (\ref{Ndef}).
In evaluating (\ref{newprob}), it is convenient to use the identity (\ref{otheridentity})
in order to perform the Kaluza-Klein summations.
We thus find
\beqn
     P_{a'\to a'}(t) &=& {1\over N^2} \biggl\lbrack
         \sum_\lambda \,\tilde \lambda^8 \,A_\lambda^4 \,e^{-\Gamma_\lambda t}\nonumber\\
          &&~~~+~ 2\, \sum_{\lambda' < \lambda}  \, 
         \tilde \lambda^4 (\tilde \lambda')^4 \, A_\lambda^2  A_{\lambda'}^2
       \, e^{-(\Gamma_\lambda + \Gamma_{\lambda'})t/2}
         \, \cos\left(  {[\lambda^2 - (\lambda')^2 ] t \over 2 p }\right)\biggr\rbrack~,~~~~~
\label{oldPresult}
\eeqn
implying a period-averaged probability
\beq
     \langle P_{a'\to a'}(t) \rangle ~=~  {1\over N^2}
         \sum_\lambda \,\tilde \lambda^8 \,A_\lambda^4\, e^{-\Gamma_\lambda t}~.
\label{oldPvevresult}
\eeq

Just as with $P_{0\to 0}(t)$, we will find it convenient to distinguish
between
two different effects:  the overall ``damping'' that arises due to axion
decays (encoded within the decay widths $\Gamma_\lambda$), and 
the oscillations that arise
due to the non-trivial mixings of the excited Kaluza-Klein states.
In order to concentrate on the latter effect, we shall 
therefore set $\Gamma_\lambda=0$ for simplicity.  Moreover, as we shall
see, this assumption will not change our phenomenological 
results {\it regardless}\/ of the time interval $t$ in (\ref{Pvevresult}).
Thus, taking $\Gamma_\lambda=0$, we 
see that (\ref{oldPresult}) reduces to
\beq
     P_{a'\to a'}(t) ~=~ {1\over N^2} \left\lbrack
         \sum_\lambda \,\tilde \lambda^8 \,A_\lambda^4 
       ~+~ 2\, \sum_{\lambda' < \lambda}  \, 
         \tilde \lambda^4 (\tilde \lambda')^4 \, A_\lambda^2  A_{\lambda'}^2
        \, \cos\left(  {[\lambda^2 - (\lambda')^2 ] t \over 2 p }\right)\right\rbrack~,
\label{Presult}
\eeq
implying a time-averaged probability
\beq
     \langle P_{a'\to a'}(t) \rangle ~=~  {1\over N^2}
         \sum_\lambda \,\tilde \lambda^8 \,A_\lambda^4~.
\label{Pvevresult}
\eeq
  
It is straightforward to evaluate this time-averaged
probability for different values of $y$.  Remarkably, however, we find that 
\beq
       \lim_{n_{\rm max}\to \infty} 
          ~ \langle P_{a'\to a'}(t) \rangle ~=~ 0 ~!
\label{limitingcase}
\eeq
We therefore have virtually no probability for detecting the 
linear combination $a'$ at any later time after it is produced!
Indeed, at the initial time $t=0$, the axion probability starts at 1.
This is guaranteed by the unitarity of the $U$-matrix in (\ref{Umatrix}),
and may be verified directly from (\ref{Presult}).
However, for $t>0$, the multi-component Kaluza-Klein oscillations
drive the net probability rapidly to zero.
Indeed, as we shall see, this holds for all $t> 0$, and does not rely   
on taking the $t\to \infty$ limit.
At no later time does a macroscopic axion regeneration appear.
It is for this reason that it is justified to set $\Gamma_\lambda=0$
in (\ref{oldPresult}) and (\ref{oldPvevresult}).

\begin{figure}[ht]
\centerline{
      \epsfxsize 4.0 truein \epsfbox {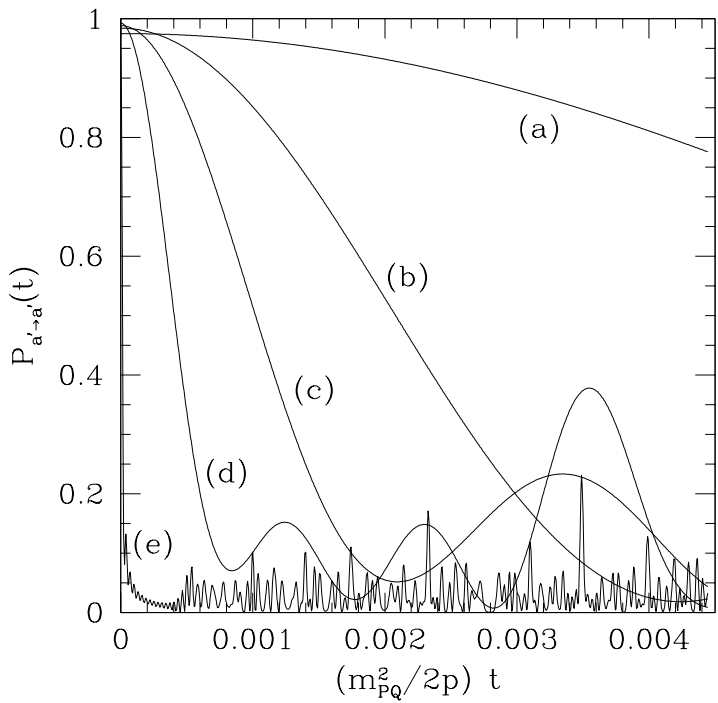}}
\caption{The axion preservation probability $P_{a'\to a'}(t)$
    as a function of the number $n_{\rm max}$ of Kaluza-Klein  
    states which are included in the system.
    For this plot we have set $y=15$, and 
    taken (a) $n_{\rm max}=1$;   (b) $n_{\rm max}=2$;   
         (c) $n_{\rm max}=3$;   (d) $n_{\rm max}=5$;  and   (e) $n_{\rm max}=30$.
    As $n_{\rm max}$ increases, the axion probability
    rapidly falls to zero as a result of the destructive
   interference of the Kaluza-Klein states, 
   and remains suppressed 
     without significant axion regeneration at any later times.
     Note, in particular, that the ``spikes'' in this plot are
     also suppressed as $\sim 1/n_{\rm max}$, and vanish for large
     $n_{\rm max}$.
    Thus, for significantly large $n_{\rm max}$, the destructive
    interference of the excited Kaluza-Klein states 
    causes the axion to ``decohere'', implying that there is
    negligible probability for subsequently detecting 
     the original axion state at any future time.} 
\label{dropping}
\end{figure}

Of course, these conclusions rely on taking the $n_{\rm max}\to \infty$ limit
in (\ref{limitingcase}).
However, even if we truncate $n_{\rm max}$ at a finite value 
$\sim {\cal O}(M_{\rm string} R)$
(reflecting the expected limit of validity for our effective field-theoretic
treatment), we find
\beq
           \langle P_{a'\to a'}(t) \rangle ~\lsim ~ (M_{\rm string} R)^{-1}~.
\label{probvolume}
\eeq
For $M_{\rm string}\approx 10$ TeV  and $R\approx 1$ millimeter,
this implies  
\beq
           \langle P_{a'\to a'}(t) \rangle ~\lsim~ {\cal O}(10^{-17})~.
\eeq
Such an axion-mediated process therefore continues to be truly invisible in the
sense that the $a'$ state literally ``disappears'' over a very short time interval,
with only a vanishingly small probability for detection of the $a'$ state
at any later time. 
This behavior is illustrated in Fig.~\ref{dropping}.
As we have noted, this is important because
the $a'$ state couples to Standard-Model fields 
with an unsuppressed coupling $1/f_{\rm PQ}$ rather than
with a volume-suppressed coupling $1/\hat f_{\rm PQ}$.

Note that as $y\to \infty$ (reproducing the four-dimensional
limit), the time needed for the probability to drop to zero 
increases without bound.  Thus, in the four-dimensional limit $y\to \infty$,
the probability begins and remains at 1, as expected.
However, for all finite values of $y$, the axion probability $P_{a'\to a'}(t)$
drops to zero in finite time, and remains there (as shown in Fig.~\ref{dropping}). 
Indeed, for $n_{\rm max}\gg 1$, we see that
the time $\tau_0$ for the axion probability to drop to 
a predetermined fraction of its initial value scales as\footnote{
      Note that the derivations of these scaling results 
      can often be quite subtle.
      In order to derive (\ref{droppingtime}), 
      we first observe that $\tau_0 \sim  t_0/N^2 $, 
      where $N$ is the normalization factor in (\ref{Ndef}) 
      and where $t_0$ is a time scale that is independent of $n_{\rm max}$. 
      This scaling behavior holds independently of $y$ 
         when $n_{\rm max}\gg 1$, as can be verified numerically.  
       Given this, it is straightforward
      to investigate the behavior of $t_0$ as a function of $y$, leading
      to the result $t_0\sim y^{-2}$.  We thus obtain $\tau_0\sim 
       (n_{\rm max} y)^{-2}$.  The final step is to realize that
      there is a hidden $y$-dependence buried in the meaning of the 
       cutoff $n_{\rm max}$, and that in order to compare cutoffs for
      different values of $y$, we must choose a uniform $y$-independent
      convention for the Kaluza-Klein truncation.
      Specifically, for each value of $y$, 
      we must choose an appropriate $y$-dependent normalization of 
       the cutoff $n_{\rm max}$ such that 
       the rescaled cutoff $n'_{\rm max}$ has a fixed,
       $y$-independent net effect in an eigenvalue sum such as (\ref{Presult}). 
      It turns out that such a renormalization
       compels us to choose $n'_{\rm max} \sim 
        y^2 n_{\rm max}$.
       We therefore find
       $\tau_0 \sim y^2/(n'_{\rm max})^2$.
        To provide some explicit numbers, 
         let us define $\tau_0$ as the time needed for the axion
        probability to fall to $10\%$ of its initial value, and let us 
       likewise define $n_{\rm crit}$ to be the minimum number of Kaluza-Klein eigenvalues
        that must be included in the sum in (\ref{Presult}) order to produce 
         an initial axion probability of 0.99.  
      (In this connection, note that it is only in the formal $n_{\rm max}\to\infty$
      limit that the initial probability truly approaches 1.)
         We then find that 
       $t_0 \approx 9.665 / y^2$ and $n_{\rm crit} \approx 981/y^2$.
         This latter relation enables us to normalize our values of
          $n_{\rm max}$ in relation to $n_{\rm crit}$ by defining
          $n'_{\rm max}\equiv n_{\rm max}/ n_{\rm crit} =  
         y^2 n_{\rm max}/981$, leading to the final result
   \beq
      \left({m_{\rm PQ}^2 \over 2p}\right) \tau_0 ~\approx~  10^{-5}\, 
            {y^2\over ({n'_{\rm max}})^2 } ~.
   \label{decoherencetime}
   \eeq
         The value of the rescaled cutoff $n'_{\rm max}$ is 
         then arbitrary, and may be chosen according to 
         considerations beyond those of our effective field-theory approach
         (such as truncating according to the underlying string scale). 
    } 
\beq
            \tau_0 ~\sim~ {y^2\over n_{\rm max}^2}~.
\label{droppingtime}
\eeq
Thus, as $y\to \infty$, we see that $\tau_0 \to\infty$, while
for finite values of $y$ this ``decoherence time'' is extremely short.
This results in an essentially immediate suppression of the axion
probability.
Moreover, as we shall verify later in this section, this decoherence time
is substantially smaller than the lifetime of the heaviest Kaluza-Klein
mode contained within $a'$.
This decoherence mechanism therefore renders the
axion superposition $a'$ virtually invisible with respect to subsequent 
axion interactions involving $a'$.

It is straightforward to understand this suppression at an intuitive level.
Unlike the case with the simple axion zero-mode $a_0$, in the present
case our initial axion state is the infinite linear superposition 
$a'$ given in (\ref{apdef}).
Let us therefore consider the behavior of the individual terms in the 
probability sum (\ref{Presult}) as $n_{\rm max}$ gets large.
For this purpose we may drop
all factors of two and focus only on the behavior
of $P_{a'\to a'}$ as a function of $n_{\rm max}$. 
For fixed $y$ and large eigenvalues $\tilde \lambda$, we have
$\tilde \lambda^4  A_\lambda^2 \approx  2$.
Therefore, as we introduce increasingly heavy eigenvalues into the probability
sum (\ref{Presult}), we find the effective behavior
\beq
     P_{a'\to a'}(t) ~\sim~ {1\over n_{\rm max}^2} \,
      \left\lbrack
         \sum_\lambda \,(1) ~+~
         2\, \sum_{\lambda' < \lambda}  (1) \, 
            \, \cos\left(  
         {[\lambda^2 - (\lambda')^2 ] t \over 2 p }\right)\right\rbrack~
\label{effectivebehavior}
\eeq
where each $\lambda$-sum contains $n_{\rm max}$ terms.
For $t=0$, this result factorizes to take the simple leading form
\beq
     P_{a'\to a'}(0) ~\sim~ {1\over n_{\rm max}^2} \,
         \sum_{\lambda,\lambda'} \,(1+1) ~
     \sim~ {1\over n_{\rm max}^2} \,
         \cdot \,n_{\rm max}^2~. 
\eeq
In other words, the presence of two independent $\lambda$-sums provides
an effective factor of $n_{\rm max}^2$ which cancels the factor of
$n_{\rm max}^{-2}$ that resulted from the normalization of $a'$.
This initial coherent contribution of the independent two $\lambda$-sums 
enables the initial probability to start at 1 regardless
of the size of the volume of the compactified space.
Indeed, the initial state $a'$ is a highly coherent state.
However, at later times $t>0$, the cosine terms in (\ref{effectivebehavior})
no longer add coherently to the sum, and their destructive interference
effectively causes the sum to scale only as $n_{\rm max}$,
corresponding to a single diagonal $\lambda$-sum:
\beq
     P_{a'\to a'}(t) ~\sim~ {1\over n_{\rm max}^2} \,
         \sum_{\lambda} \,(1) ~\sim~ 
     {1\over n_{\rm max}^2} \,\cdot\, 
         n_{\rm max} ~\sim~ {1\over n_{\rm max}}~.
\label{eqnohere}
\eeq
This phenomenon is completely
analogous to the fact that a random walk traverses only the square root
of the distance traversed by a coherent, directed walk.  
It is for this reason that
the net axion preservation probability $P_{a'\to a'}(t)$ is so
strikingly suppressed.
Essentially, the initial axion state $a'$ has ``decohered'' as a result
of the incoherent Kaluza-Klein oscillations induced by the non-diagonal axion
mass matrix.

We thus conclude that in higher dimensions, 
all axion-mediated processes
which rely on the production and subsequent detection of the $a'$ mode are
strongly suppressed by a rapid ``decoherence'' which
renders them virtually ``invisible''.
This decoherence arises as a result of the destructive 
interference of the infinite-component laboratory
axion oscillations.
We see, then, that this provides an entirely new higher-dimensional
mechanism which can contribute to the invisibility of 
$a'$-mediated processes. 

It is important to stress that this decoherence mechanism is relevant 
only for those processes which are sensitive to the time-evolution 
of the axion.
As we have indicated, this includes all axion-mediated process
(\eg, axion-exchange processes), as well as 
processes in which the axion is directly detected in the laboratory.
Moreover, measurements of the axion flux from the Sun or from supernovae
also fall into this category.
However, this does {\it not}\/ include processes which 
are insensitive to axion time-evolution.  
These include, for example, axion-production processes
in which the axion appears only as missing energy.
Nevertheless, invisibility can be achieved for such processes
by adjusting the value of $\hat f_{\rm PQ}$ via the mechanisms
discussed in Sects.~2 and 4.

Finally, we 
remark that our higher-dimensional decoherence mechanism is 
completely general, and applies not only to axions, but also to any bulk 
field $\phi$ whose ``shadow'' interaction with the Standard-Model brane
involves a coupling between Standard-Model fields 
and a ``brane shadow'' Kaluza-Klein superposition
\beq
     \phi(y=0) ~=~ \sum_k \,c_k \,\phi_k  
\eeq
with non-zero coefficients $c_k$.
Just as in the axion case considered above,  
the non-trivial time-evolution of the Kaluza-Klein modes
will cause the initial superposition $\phi(y=0)$ to ``decohere''
extremely rapidly.
This will necessarily produce a severe damping
of any Standard-Model process on the brane that involves couplings
to the coherent state $\phi(y=0)$.
For example,
this might therefore provide a partial solution to 
the notorious dilaton problem in string theory:  it may simply be
that the string-theoretic dilaton is ``invisible'', in much the
same way as the axion is invisible.
Similar considerations may also apply to Kaluza-Klein gravitons
as well as other bulk moduli fields.

\subsection{Laboratory oscillations inducing $a_k \to a'$}

Finally, let us consider a third relevant oscillation probability.  
As we discussed above, fields on the Standard-Model brane can couple
only to the linear combination $a'$.  This is why the probability
$P_{a'\to a'}(t)$ is the appropriate probability for processes
involving both Standard-Model production and detection of axions.
However, for axions that are produced through mechanisms involving bulk
fields (which are non-localized) rather than brane fields (which are
localized), it is possible to envisage situations in which a single
Kaluza-Klein mode $a_k$ (\eg, the zero-mode $a_0$) is produced.  
However, detection of such an axion mode on our Standard-Model brane 
continues to involve couplings to $a'$, and therefore in 
such cases a relevant probability for detection is given by
\beqn
         P_{a_0\to a'}(t) &=& 
     {1\over N} \left| \sum_{\ell=0}^\infty r_\ell A_{0\to\ell}(t)\right|^2\nonumber\\
     &=&  {1\over N} \biggl\lbrack
         \sum_\lambda \,\tilde \lambda^4 \,A_\lambda^4 \,
         e^{-\Gamma_\lambda t}~+~ \nonumber\\
   && ~~~~
         2\, \sum_{\lambda' < \lambda}  \, 
         \tilde \lambda^2 (\tilde \lambda')^2 \, A_\lambda^2  A_{\lambda'}^2
         \,e^{-(\Gamma_\lambda + \Gamma_{\lambda'})t/2}
         \, \cos\left(  {[\lambda^2 - (\lambda')^2 ] t \over 2 p }\right)\biggr\rbrack~.~~~~~
\eeqn
However, just as in the previous case, this probability vanishes in the
limit $n_{\rm max}\to\infty$, even if we again set $\Gamma_\lambda=0$.
Specifically, the probability to produce the specific coherent
state $a'$ on the Standard-Model brane vanishes 
as $n_{\rm max}\to\infty$.
Moreover, the same is true for all probabilities
$P_{a_k\to a'}(t)$ for all $k$ and for all $y$.  
Of course, we cannot interpret this result as
a decoherence, since even the initial probability at $t=0$ vanishes as
$n_{\rm max}\to\infty$ due to the negligible overlap between $a_k$ and $a'$.
Nevertheless, this demonstrates that even if
the axion is produced by bulk
fields not restricted to the Standard-Model brane, the probability for
its subsequent detection in the laboratory is vanishingly small.

\subsection{Axion detection and decay}

Finally,  we shall conclude this section by discussing  
a number of additional effects that
arise due to the existence of an infinite tower
of excited Kaluza-Klein axion states.
Our comments in this subsection will primarily be focused on the
possibility of axion detection and in particular on the role of 
axion decays. 

First, as already mentioned above, we must distinguish between 
processes in which the time-evolution of the axion plays
a role and processes which are insensitive to the time-evolution of the axion.  
For example, the latter include axion-production processes in which the axion 
is emitted into the bulk and therefore is manifested on the Standard-Model brane
only as missing energy.
A simple example of this is the axion-emission process
\beq
            F\tilde F ~\to ~a'~ \to ~{\rm bulk}~ 
\eeq
where the axion, once produced, flies into the bulk.
As we have seen in Sect.~2, this process scales as $1/f_{\rm PQ}$ rather
than $1/\hat f_{\rm PQ}$.  This already leads to the severe constraint 
$f_{\rm PQ}\gsim {\cal O}$(TeV), and is completely analogous to the 
possibilities for detecting graviton emission in 
upcoming TeV-scale collider experiments.
We shall discuss one such missing-energy signature in more detail
in Sect.~5.

By contrast, processes which involve an actual {\it detection}\/ of the
axion on the Standard-Model brane are necessarily {\it time-dependent}\/
because they involve a non-zero time interval  
between axion production and axion detection.
However, even within this category of time-dependent processes, 
there are further subdivisions that can be made.  
One important distinction
is the length of the time interval between axion production and 
axion detection.  
Certain axion-mediated processes, such as those taking place entirely within
accelerator experiments, take place on time scales that are very short
compared to the axion lifetime.  By contrast, others (such as those 
involving fluxes of axions which are 
produced in the Sun or in supernovae and which are subsequently detected on 
Earth) involve much longer time scales.  

This distinction between ``long'' and ``short'' time scales 
therefore depends on the axion lifetime.
Ordinarily, in four dimensions, the axion is relatively long-lived
because it is so light and so weakly coupled to ordinary matter. 
Indeed, the dominant axion decay mode is to two photons, yielding 
the lifetime 
\beq
           \tau_{\rm 4D}(a\to\gamma\gamma) ~=~ 
          {1\over \Gamma_{\rm 4D}(a\to \gamma\gamma)} ~\approx~
               \left({4\pi \over \alpha}\right)^2 
         \left({ \hat f_{\rm PQ}^2\over m_{\rm PQ}^3} \right)~\approx~
                10^{48}~{\rm seconds}~.
\eeq
In this expression, we are 
writing $\hat f_{\rm PQ}\approx 10^{12}$~GeV to denote the usual 
four-dimensional Peccei-Quinn symmetry-breaking scale,
and taking $m_{\rm PQ}\approx 10^{-5}$~eV to denote the usual 
four-dimensional axion mass.
(We choose these symbols in order to facilitate the comparison between
the four-dimensional and higher-dimensional situations.)
We are also disregarding numerous model-dependent ${\cal O}(1)$
coefficients which do not affect the overall scale of the result.
Thus, in four dimensions, the axion is extraordinarily stable. 

In higher dimensions, this situation changes dramatically.
Of course, the coherent axion mode $a'$ is not a mass eigenstate,
and thus, strictly speaking, it does not have a well-defined ``lifetime''.
Nevertheless, 
we can determine an effective lifetime for $a'$ by estimating the
shortest lifetime of any of the mass eigenstates $\hat a_\lambda$ 
of which it is comprised.
This then yields the time scale over which the 
coherent state $a'$ naturally decays as a result of its couplings to
ordinary matter.  In general, it is the heaviest Kaluza-Klein  
mass eigenstates which have the shortest lifetimes, with the decay
mode into two photons continuing to be dominant.
As a function of the cutoff $n_{\rm max}$,   
the ``lifetime'' of the coherent state $a'$ can therefore be
estimated to be
\beqn
    \tau_{D>4}(a'\to \gamma\gamma)  &=& 
          {1\over \Gamma_{\rm 4D}(\hat a_{\lambda_{\rm max}}\to \gamma\gamma)} 
       ~\approx~ 
          \left({4\pi \over \alpha}\right)^2 
         \left({ \hat f_{\rm PQ}^2\over M_{\rm string}^3} \right)~\nonumber\\
          &=& \left( {m_{\rm PQ}\over M_{\rm string}}\right)^3 
       \tau_{\rm 4D}(a\to \gamma\gamma) ~\approx~ 10^{-3}~{\rm seconds}~.
\label{lifetime}
\eeqn
In this expression we have taken $n_{\rm max}\approx R M_{\rm string}$,
implying $\lambda_{\rm max}\approx M_{\rm string}$.  We have also
chosen $M_{\rm string}=1$~TeV for simplicity.
Thus, we see that the coherent state can be expected to decay
to two photons much more rapidly than the usual four-dimensional axion.
Of course, in this calculation we have taken the cutoff 
$\lambda_{\rm max}\approx M_{\rm string}$, which represents the ``worst-case''
scenario.  In any axion-related process of total energy $E<M_{\rm string}$, 
the production of Kaluza-Klein axion modes $\hat a_\lambda$ 
with $\lambda\gsim E$ will be kinematically disfavored.
We would then take $\lambda_{\rm max}\approx E$,
which can increase the lifetime considerably.

 {\it A priori}\/, this significant reduction in the axion lifetime 
relative to the four-dimensional case
means that our ``decoherence'' phenomenon for $a'$ is 
irrelevant unless the decoherence time $\tau_0$ is even smaller.
However, it is straightforward to verify that this is indeed the case.
Consulting (\ref{decoherencetime}), we see that 
\beq
          \tau_0~\approx~ y^4 \, \left( {p\over 1~{\rm TeV}}\right)\,
              \left({M_{\rm string}\over 1~{\rm TeV}}\right)^{-2} \,\times\,
                10^{-33}~{\rm seconds}~
\label{decohtime}
\eeq
where we have taken $n'_{\rm max}\approx RM_{\rm string}$.
For all phenomenologically interesting values of $p$ and $M_{\rm string}$,
we thus conclude that $\tau_0 \ll \tau_{D>4}$.  We see, therefore, that
the coherent axion mode $a'$
indeed decoheres sufficiently rapidly to justify the 
$\Gamma_\Lambda=0$ approximation that was used in deriving 
it.\footnote{In comparing (\ref{lifetime}) and (\ref{decohtime}), we
      must actually account for the relative rescaling of the cutoff
      $n_{\rm max}\to n'_{\rm max}$, as discussed 
      above (\ref{decoherencetime}).
      This introduces an additional multiplicative factor $(y^2/981)^3$ 
      into (\ref{lifetime}), thereby shortening $\tau_{D>4}$ by an additional
      factor $\approx 10^{-9}$ for $y\approx {\cal O}(1)$.
      However, we see from (\ref{decohtime}) that this still does not
      affect our main conclusion that $\tau_0\ll \tau_{D>4}$.}

Given that the coherent state $a'$ has an
intrinsic  lifetime $\tau_{a'}\approx 10^{-3}$~seconds, 
we can therefore use this as a benchmark  
for separating ``long'' and ``short'' processes.
For laboratory oscillations over time scales shorter than this,
it is legitimate to neglect the axion
decay widths in calculations of these oscillations.
By contrast, calculations of oscillations over longer time scales  
require the inclusion of the decay widths, and will therefore start to feel the
effects of axion decays.

It is therefore important
to understand the effects of such axion decays, particularly as they
relate to axion ``invisibility''.
For this purpose, we may draw another distinction, this time between
processes
that take place entirely within a single detector (such as 
an axion-mediated process $F\tilde F\to a' \to F\tilde F$), and those which
take place largely outside our detector (such as an axion beam
travelling from the Sun to the Earth).  
For processes taking place entirely within 
a single detector, axion decay represents a breakdown of invisibility
because we can in principle detect the emitted decay products.
For example, even though we have found that $P_{a'\to a'}(t)$ is suppressed
for all times $t$ exceeding the decoherence time $\tau_0$, the decohered
state will nevertheless continue to propagate until 
the individual Kaluza-Klein axion modes that comprise this
decohered state themselves decay.
This will be discussed in more detail below.
Given this observation, one 
might initially doubt the phenomenological importance
of the $a'$ decoherence phenomenon.
However, the important point is that it is only the
 {\it coherent}\/ state $a'$ which couples to Standard-Model fields with
the potentially dangerous 
unsuppressed coupling $1/f_{\rm PQ}$, whereas individual Kaluza-Klein
axion modes instead experience the safer suppressed coupling 
$1/\hat f_{\rm PQ}$.
The decoherence phenomenon therefore indicates that there is only
a vanishingly small time interval during which a process 
of the form $F\tilde F\to a'\to F\tilde F$  can possibly occur
with a dangerously large amplitude scaling as $1/f_{\rm PQ}^2$.
After this initial time interval, the $a'$ state decoheres,
and all subsequent interactions and decays will have amplitudes
scaling as $1/\hat f_{\rm PQ}^2$. 

In sharp contrast are processes which largely
take place {\it outside}\/ our detectors.
In such cases, we do not expect to be able
to detect the decay products that result from axion decays at intermediate times.
For example, in the case of an axion beam travelling
from the Sun to the Earth, the photons emitted in flight through axion decays 
are presumably indistinguishable from the general background radiation.
In such situations, therefore,
axion decays lead not to a loss of invisibility, but rather to an
  {\it enhancement}\/ of it.
For example, let us consider the probability 
that the initial state $a'$ will be found in a particular mass eigenstate
$\hat a_\lambda$ as function of time.  Following the same procedure as
above and using (\ref{apdeflambda}), we find the probability
\beq
           P_{a'\to \hat a_\lambda}(t) ~=~ {1\over N}\,\tilde\lambda^4 \,
                     A_\lambda^2 \, e^{-\Gamma_\lambda t}~.
\label{aptolambda}
\eeq
Note that this is an exact result valid for all times, with axion decays 
producing an exponential suppression for the probability.
Moreover, as we shall now demonstrate, this fact can be used to 
provide a direct experimental test 
of the higher-dimensional nature of the axion.
Using (\ref{aptolambda}), we may define
\beq
        P_{\rm tot}(t) ~\equiv~  \sum_\lambda P_{a'\to \hat a_\lambda}~=~ 
                {1\over N} \sum_\lambda \tilde\lambda^4 \,
                     A_\lambda^2 \, e^{-\Gamma_\lambda t}~
\eeq
as a ``collective'' amplitude that the axion mode $a'$ survives
as a function of time.  This interpretation is justified because
$P_{\rm tot}(t)$ is nothing but the time-dependent norm 
of the original $a'$ superposition:
\beq
          P_{\rm tot}(t) ~=~ \langle a'(t)|a'(t)\rangle~.
\eeq
Given this norm, the collective decay width 
(\ie, the instantaneous decay probability per unit time) is given as
\beq
       \langle \Gamma \rangle ~\equiv~ 
        - {1\over P_{\rm tot}} {dP_{\rm tot}(t)\over dt} ~=~
          {\sum_\lambda \,\Gamma_\lambda \,\tilde\lambda^4 \,
            A_\lambda^2 \, e^{-\Gamma_\lambda t} \over
          \sum_\lambda \,\tilde\lambda^4 \, A_\lambda^2 \, e^{-\Gamma_\lambda t}}~,
\eeq
in complete analogy with the formalism for radioactive decays.
However, because of the presence of an infinite tower of Kaluza-Klein states,
it is immediately apparent that $\langle \Gamma \rangle$ 
is itself a function of time!
For example, at very early times we have 
$\langle \Gamma\rangle \sim \Gamma_{\lambda_{\rm max}}\sim 
M_{\rm string}^3/\hat f_{\rm PQ}^2$,
where $\lambda_{\rm max}\approx M_{\rm string}$ is 
the heaviest mass eigenvalue included in the linear superposition.
By contrast, at extremely late times, we have $\langle \Gamma\rangle 
\to \Gamma_{\lambda_0}$ where $\lambda_0$ is the {\it lightest}\/ mass 
eigenvalue.  Since $\langle \Gamma\rangle$ is the instantaneous decay probability 
per unit time (which can be measured in our Earth-bound detector),  
a time-variation in $\langle \Gamma \rangle$ from different axion sources
at different distances would serve as a direct experimental test of
the higher-dimensional nature of the axion.

Finally, let us briefly comment on the possibility of measuring the
laboratory axion oscillations discussed previously.
Although we have shown that probabilities such as 
$P_{a_k\to a_\ell}(t)$ 
experience sinusoidal oscillations as functions of time, at a physical level
these calculations presuppose that we are capable of experimentally
detecting individual Kaluza-Klein modes $a_k$.   
In other words, these calculations presuppose that we are able to
distinguish the final axion states according to their Kaluza-Klein
quantum numbers.
In situations where such detections can be made, our previous results
continue to apply.
However, in the most straightforward scenarios, all Kaluza-Klein states
will have identical decay modes (\eg, into two photons), leaving us
with no experimental ``handle'' through which to detect the presence of 
an oscillation.
Thus, the prospects for detecting laboratory axion oscillations depend
crucially on the ability to perform laboratory axion measurements which 
are sensitive to particular Kaluza-Klein quantum numbers.
Of course, this is completely analogous to the case of neutrino oscillations,
where the existence of neutrino decay modes that distinguish
between different neutrino quantum numbers 
(such as $SU(2)$ gauge charge or flavor)
permit the detection of neutrino oscillations.

\section{Cosmological relic axion oscillations}
\setcounter{footnote}{0}

In this section, we shall discuss a third higher-dimensional
effect that can contribute to the ``invisibility'' of axions:  the
rate at which the energy trapped in cosmological relic axion 
oscillations is dissipated.   
As we shall see, the presence of an infinite tower of Kaluza-Klein
axion states can, under certain circumstances, actually {\it enhance}\/
the rate at which this oscillation energy is dissipated.
In such cases, 
the effective Peccei-Quinn scale $\hat f_{\rm PQ}$ 
can therefore be raised beyond its usual relic-oscillation bounds,
leading to an axion whose couplings to matter are even more suppressed
than in four dimensions.  This can therefore provide another 
higher-dimensional way of achieving an ``invisible'' axion.

Let us first recall the situation in four dimensions.
One of the most important constraints on the scale of Peccei-Quinn
symmetry breaking, and hence on the mass of the axion and the strength
of its couplings to ordinary matter, comes from
cosmological relic axion oscillations.  Unlike the laboratory
oscillations discussed in the previous section, these cosmological
relic oscillations arise due to the fact that 
as the universe cools and passes through the QCD phase transition
at $T\approx \Lambda_{\rm QCD}$,
instanton effects suddenly establish a non-zero axion potential
where none previously existed.  In the usual four-dimensional
situation, the axion can therefore find
itself displaced relative to the newly-established minimum
of the potential, and begin to oscillate around it according to the
differential equation
\beq
     { d^2 a \over dt^2} ~+~ 3H(t)\,{da\over dt} ~+~ m_a^2\, a~=~0~,
                 ~~~~~~~~ t\gsim t_{\rm QCD}~.
\label{4ddiffeq}
\eeq
Here $H(t)=1/(2t)$ is the Hubble constant,
where we are assuming for simplicity 
that this time-evolution takes place entirely in a radiation-dominated universe. 
We are also assuming that the axion mass $m_a$ is
independent of time for $t\gsim t_{\rm QCD}$ and vanishes for times
$t\lsim t_{\rm QCD}$.
We are also neglecting the decay width of the axion.
In general, the initial amplitude of this oscillation at $t_{\rm QCD}$ is 
set by the initial random angular displacement of the axion field. 
This scales as $f_{\rm PQ}$, where $f_{\rm PQ}$ is the axion decay constant. 
Although these oscillations are ultimately damped 
due the cosmological Hubble expansion term,  a relic of these axion oscillations
should still exist today.   Imposing the requirement that the energy stored
in the relic oscillation today be less than the critical energy
density (so as not to overclose the universe)
then sets an upper bound $f_{\rm PQ}\lsim 10^{12}$ GeV which is consistent
with bounds obtained through other means.
This is therefore a bound on the ``invisibility'' of the axion.

At first glance, it might seem that this issue should no longer
play a role in our higher-dimensional scenario in which the fundamental
scale $f_{\rm PQ}$ of Peccei-Quinn symmetry breaking is substantially 
lowered (perhaps even to the TeV-range) as a result of the volume
factor in (\ref{fhatdef}).  Indeed, it is the (low) fundamental mass scale $f_{\rm PQ}$ 
rather than the (high) effective mass scale $\hat f_{\rm PQ}$
which sets the size of the VEV of the axion field $a$ in 
the five-dimensional Lagrangian (\ref{lagfive}), and which similarly sets 
the overall scale of the initial random angular displacement of the axion field. 
However, in passing from (\ref{lagfive}) to the four-dimensional Lagrangian
(\ref{lag}), there is an implicit volume-dependent
rescaling of the axion field, as discussed below (\ref{rdefs}).
Thus, even though initial displacement of the unrescaled 
axion field scales with $f_{\rm PQ}$,
the initial displacement of the rescaled axion field scales with $\hat f_{\rm PQ}$.
The danger of an excessive oscillation energy density today 
is therefore just as relevant 
for our higher-dimensional scenario as it is for the usual four-dimensional case.

\begin{figure}[ht]
\centerline{
      \epsfxsize 3.7 truein \epsfbox {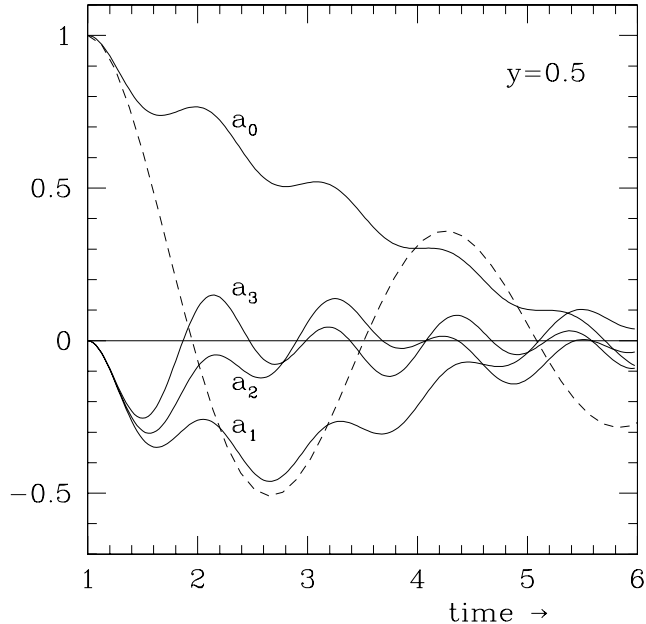}}
\caption{ A plot of the coupled Kaluza-Klein
           cosmological relic axion oscillations.
          For this plot, we have normalized the initial
           displacement of the axion zero-mode to $1$, and
           taken $y=0.5$, $m_{\rm PQ}=2$, and $t_{\rm QCD}=1$ 
         (in dimensionless units).  We have also considered the effects 
         of only the first three excited Kaluza-Klein modes.
         Although the excited Kaluza-Klein modes have 
         vanishing initial displacements, they are triggered into
          oscillation as a result of the initial displacement 
          of the zero-mode.
          This in turn changes the subsequent time-evolution
            of the zero-mode. 
         By contrast, the superimposed dashed line shows the behavior 
          of the usual four-dimensional axion zero-mode  
           in the $y\to\infty$ limit (when no Kaluza-Klein modes are present).}
\label{kkosc}
\end{figure}

In fact, this danger may actually be greater in the higher-dimensional
case.  This is because the Kaluza-Klein reduction yields not only the usual
zero-mode axion $a_0$, but also an infinite tower of excited Kaluza-Klein axions 
$a_k$ ($k>0$).
The above differential equation for the axion oscillations 
then generalizes to 
\beq
     { d^2 a_k \over dt^2} ~+~ 3H(t)\,{da_k\over dt} ~+~ 
        {\cal M}^2_{k\ell} \, a_\ell~=~0~,
                 ~~~~~~~~ t\gsim t_{\rm QCD}~
\label{newdiffeq}
\eeq
where ${\cal M}^2$ is the non-diagonal mass matrix given in (\ref{matrix}).
(Note that we shall continue to neglect axion decay widths in these equations;
this assumption will be justified at the end of this section.)
Of course, due to their Kaluza-Klein masses, these excited Kaluza-Klein axions
feel a non-zero potential even prior to the QCD phase transition (\ie, even prior
to the ``turn-on'' of $m_{\rm PQ}$), and it is
therefore reasonable to assume that they are already sitting at their minima
at the time of the QCD phase transition.  However, due to the non-diagonal mass
mixing matrix in (\ref{newdiffeq}), the initial displacement
of only the zero-mode axion $a_0$ is sufficient to 
trigger the excited Kaluza-Klein modes into oscillation. 
This situation is illustrated in Fig.~\ref{kkosc}.

Given this observation, there are {\it a priori}\/ three possible effects that
these excited Kaluza-Klein states can have on the system.  First, it is possible 
that these
excited Kaluza-Klein states will ``capture'' oscillation energy from the zero-mode
oscillation, and essentially store it.  Thus, in this case, the total energy
density of the system would dissipate more slowly, thereby resulting in a
greater relic energy density today.  This would seriously strengthen the five-dimensional
axion bounds relative to the usual four-dimensional bounds, and provide the most
serious threat to the viability of these higher-dimensional scenarios.
The second possibility is that although the excited Kaluza-Klein  
states steal energy from the zero-mode oscillation, they may be able to dissipate
it {\it more}\/ effectively.  
This would then lead to an {\it enhanced}\/ relic energy loss rate,
implying a weakened bound on the higher-dimensional scenarios.
Finally, the third possibility is that the two effects cancel exactly, with
the excited Kaluza-Klein states oscillating in such a way that even though
they capture some energy from the zero-mode, they also alter the time-development
of the zero-mode in an exactly compensatory manner.  
Thus the traditional four-dimensional bounds would remain unchanged.

In order to determine which of these possibilities is realized, 
we must solve the coupled differential equations (\ref{newdiffeq}).
In so doing, we shall make the following assumptions.
First, we shall assume that prior to the QCD phase transition at $t_{\rm QCD}$,
the axion zero-mode experiences no potential and therefore has an initial 
displacement set to $1$ (in units of $\hat f_{\rm PQ}$).  We shall likewise assume that
$da_0/dt=0$ at $t=t_{\rm QCD}$.  By contrast, for the excited
Kaluza-Klein modes $a_k$ ($k>0$), we shall begin with the initial conditions
$a_k = da_k/dt =0$ at $t_{\rm QCD}$.  These conditions reflect the fact that these 
excited modes have non-zero 
Kaluza-Klein masses even prior to the ``turn-on'' of $m_{\rm PQ}$ at $t_{\rm QCD}$,
and hence should have essentially settled into their minima prior to the QCD 
phase transition.  Although it is possible for the lighter
Kaluza-Klein axion modes to have small displacements as well, 
for simplicity we shall ignore this possibility in
what follows.  
Note that this assumption also enables us to start with identical initial relic 
oscillation energy densities in both the four- and five-dimensional situations,
and thereby enables us to make a direct comparison of the effects of the 
Kaluza-Klein modes on the cosmological time-evolution of the system.
Finally, in substituting the matrix (\ref{matrix}) into (\ref{newdiffeq}),
we shall assume that $m_{\rm PQ}(t)$ is given exactly by the step function
\beq
             m_{\rm PQ}(t) ~=~ m_{\rm PQ} \,\Theta(t-t_{\rm QCD})~.
\eeq
The constancy of $m_{\rm PQ}(t)$ for
$t>t_{\rm QCD}$
is often referred to as the ``adiabatic'' approximation.
We caution, however, that 
for large values of $m_{\rm PQ}$ this 
step-function approximation can differ quite substantially
from the results of a more 
careful analysis of the time/temperature-dependence of the axion
mass due to instanton effects~\cite{axiontemp}.

Given these assumptions, it turns out to be possible to solve the differential
equations (\ref{newdiffeq}) {\it analytically}\/ for all times $t\geq t_{\rm QCD}$
and for an arbitrary number of Kaluza-Klein modes.
We therefore do not need to make any further approximations
(such as the traditional separation into so-called
``overdamped'', ``underdamped'', and ``critically damped'' oscillation
phases).
The first step in our analytical solution
is to decouple the differential equations (\ref{newdiffeq})
by passing to the mass-eigenstate basis $\hat a_\lambda$ defined
in (\ref{masseig}).
In order to work with dimensionless quantities, we shall define
$\tilde a_\lambda \equiv \hat a_\lambda/\hat f_{\rm PQ}$. 
Each of our uncoupled differential equations then takes
the form
\beq     {d^2 \tilde a_\lambda\over dt^2}  ~+~ {3\over 2t} {d\tilde a_\lambda\over dt} 
        ~+~  
              \lambda^2\, \tilde a_\lambda ~=~ 0~.
\eeq
For simplicity we can recast this equation into the form
\beq     {d^2 \tilde a_\lambda \over d\tau^2}  ~+~ 
       {3\over 2\tau } {d\tilde a_\lambda \over d\tau} ~+~   
            \tilde a_\lambda ~=~ 0~
\eeq
where we have defined $\tau \equiv \lambda t$.
The most general solution to this equation is then given by
\beq    \tilde a(\tau) ~=~ \tau^{-1/4} \, 
          \left[ c_+ J_{1/4}(\tau) + c_- J_{-1/4}(\tau) \right]
\label{eqno1}
\eeq
where $J_\nu(\tau)$ are the Bessel functions of first kind.
We therefore wish to solve for the unknown constant coefficients $c_+$ and $c_-$.
To do this, we impose our initial conditions.  In the original Kaluza-Klein
basis,
these conditions are given by $a_k(\tau_0)= \hat f_{\rm PQ} \delta_{k0}$ and
$da_k(\tau_0)/d\tau =0$
where $\tau_0 \equiv \tau_{\rm QCD}$ is
the initial time at which we begin the time-evolution of our axion fields
(representing the ``turn-on'' time for $m_{\rm PQ}$).
In the dimensionless {\it mass-eigenstate}\/ 
basis, these initial conditions therefore take the form:
\beq   \tilde a_\lambda(\tau_0) = A_\lambda ~,~~~~~~~~~ 
               {d\tilde a_\lambda \over d\tau } \biggl|_{\tau=\tau_0} = 0~
\eeq
where $A_\lambda$ is defined in (\ref{Adef}). 
The assumptions underlying these initial conditions were discussed
above.  Thus, in the mass-eigenstate basis, we see that {\it each}\/ of the 
Kaluza-Klein modes begins with an initial displacement.
Solving for $c_\pm$ is then straightforward.
The first initial condition (the displacement condition) 
trivially gives the constraint
\beq          c_+ J_{1/4} + c_- J_{-1/4} ~=~ A_\lambda \,\tau_0^{1/4}~
\label{aconstraint}
\eeq
where for notational convenience any Bessel function written
without an argument is understood to be evaluated at $\tau_0$.
Using the Bessel-function identity
\beq
     {d\over d\tau} [\tau^{\pm \nu} J_\nu(\tau)] ~=~ 
           \pm \tau^{\pm \nu} J_{\nu\mp 1}(\tau)~,
\eeq
we see that the initial velocity constraint takes the form
\beq    {d\tilde a_\lambda \over d\tau}  ~=~  \tau^{-1/4} 
        \left[   -c_+ J_{5/4}(\tau) + c_- J_{-5/4}(\tau)\right]~,
\label{eqno2}
\eeq
implying $c_+ J_{5/4} = c_- J_{-5/4}$.
Together with (\ref{aconstraint}), this leads to the solutions
\beq           c_\pm ~=~  
        - {\pi\over \sqrt{2}} \,A_\lambda \,\tau_0^{5/4}\, J_{\mp 5/4}~
\label{cpm}
\eeq 
where we have used the further identity
\beq
           J_{1/4}J_{-5/4} + J_{-1/4}J_{5/4}  
            ~=~ -{\sqrt{2}\over \pi \tau_0}~.
\label{eqnoxx}
\eeq

Substituting $c_\pm$ from (\ref{cpm}) into the solution given in (\ref{eqno1}), 
we thus obtain our final closed-form solution to the axion differential equation:
\beq       \tilde a_\lambda (\tau) ~=~  -{\pi \over \sqrt{2}} \,A_\lambda\,  
         \tau_0^{5/4} \tau^{-1/4}\,   
       j(\tau_0;\tau)
\eeq
where we have defined
\beq     j(\tau_0;\tau) ~\equiv~ J_{-5/4}(\tau_0) J_{1/4}(\tau)   
       + J_{5/4}(\tau_0) J_{-1/4}(\tau) ~.
\eeq
This implies that the first time-derivative is given by
\beq   {d\tilde a_\lambda \over d\tau}  ~=~ 
     {\pi \over \sqrt{2}} \,A_\lambda \,  
         \tau_0^{5/4} \tau^{-1/4}\,   
          j'(\tau_0;\tau) 
\eeq
where we have likewise defined
\beq     j'(\tau_0;\tau) ~\equiv~ 
        J_{-5/4}(\tau_0) J_{5/4}(\tau)   
           - J_{5/4}(\tau_0) J_{-5/4}(\tau ) ~.
\eeq
Note that $j'(\tau_0;\tau)\to 0$ as $\tau \to \tau_0$, as expected,
since the initial velocities vanish for each of the axion modes.

Now, the energy contribution from a single mode $\tilde a_\lambda$ is given by
\beq   \tilde \rho_\lambda(\tau) ~\equiv~ {\tilde \lambda^2\over 2} 
         \left[ \tilde a_\lambda ^2 + 
        \left({d\tilde a_\lambda\over d\tau}\right)^2 \right]~
\eeq
where $\tilde \lambda$ is the dimensionless eigenvalue defined in (\ref{dimeig})
and where $\tilde \rho\equiv \rho/(m_{\rm PQ}^2 \hat f_{\rm PQ}^2)$ 
is a dimensionless energy density.
Substituting in the above results, we therefore find
\beq
     \tilde \rho_\lambda(\tau) ~=~
   {\pi^2 \over 4}\,A_\lambda^2\, \tilde\lambda^2\, 
       \tau_0^{5/2} \tau^{-1/2}\, 
   \left[ j(\tau_0;\tau)^2 + j'(\tau_0;\tau)^2 \right]~.
\label{ex}
\eeq

While this expression is exact for all times $\tau$,
it is also useful to have an approximation valid for extremely late 
times satisfying $\tau \gg 1$ (such as the present cosmological time). 
Using the Bessel-function asymptotic expansion
\beq     J_\nu(\tau) ~\approx~  \sqrt{2\over \pi \tau}~
           \cos\left(\tau - {\pi\nu \over 2} - {\pi\over 4}\right)~~~~~~ 
              {\rm as}~~~ \tau \to \infty~,
\eeq
it is straightforward to show that
\beq    j(\tau_0;\tau)^2 + j'(\tau_0;\tau)^2 ~\approx~
        {2\over \pi \tau}\, 
     \left\lbrace [J_{5/4}(\tau_0)]^2 + [J_{-5/4}(\tau_0)]^2  +  
       \sqrt{2}\, J_{5/4}(\tau_0) J_{-5/4}(\tau_0) \right\rbrace~. 
\eeq

Given the exact result (\ref{ex}),
we can convert from $\tau$ back to our original time variable $t$ 
to obtain the final closed-form solution
\beq  
    \tilde \rho_\lambda(\tilde t) ~=~
   {\pi^2 \over 4}\,\,A_\lambda^2\, \tilde \lambda^4 \, \tilde t_0^{\,5/2} \,
  \tilde t^{-1/2}\, 
   \left[ j(\tilde \lambda \tilde t_0;\tilde \lambda \tilde t)^2 + 
        j'(\tilde \lambda \tilde t_0; \tilde \lambda \tilde t)^2 \right]~
\eeq
where we have defined the dimensionless time $\tilde t\equiv m_{\rm PQ}t$.
Thus, adding together the contributions from all the mass-eigenstate modes, we obtain
the final energy density:
\beq    
     \tilde \rho(\tilde t) ~=~ \sum_\lambda \tilde \rho_\lambda(\tilde t) ~=~
   {\pi^2 \over 4}\,\, \tilde t_0^{\,5/2} \,\tilde t^{-1/2}\, 
   \sum_\lambda A_\lambda^2\, \tilde \lambda^4 \, 
   \left[ j(\tilde \lambda \tilde t_0;\tilde \lambda \tilde t)^2 + 
        j'(\tilde \lambda \tilde t_0;\tilde \lambda \tilde t)^2 \right]~.
\label{exactresult}
\eeq
Note that this is the {\it exact}\/ result for the relic oscillation
energy density as a function of time.

It is easy to verify that $\tilde \rho(\tilde t)$ has the correct
limit as $\tilde t\to \tilde t_0$.
Indeed, as $\tilde t\to \tilde t_0$, we find using the identity (\ref{eqnoxx})
that $\tilde \rho (t)\to  \tilde \rho_0$, where
\beq
        \tilde \rho_0 ~\equiv ~ {1\over 2} \sum_\lambda  
          \,\tilde \lambda^2 \, A_\lambda^2~.
\label{compar}
\eeq
This is indeed the correct value of the initial energy in the mass-eigenstate
basis, since each mass eigenmode starts with zero velocity and with initial 
displacement $A_\lambda$.
However, in our original Kaluza-Klein basis, our initial conditions at $\tilde t_0$
consist of having only the zero-mode displaced by $1$.
This implies that $\tilde \rho=1/2$.
Comparing this result with (\ref{compar}) then yields the
identity quoted in (\ref{identity}).
In particular, this identity holds for all $y$,
as can be verified directly by substituting the values of $\lambda$ and $A_\lambda$
and evaluating the eigenvalue sum.

For late times $\tilde \lambda \tilde t\gg 1$, the exact result (\ref{exactresult}) 
for the energy density simplifies to take the form
\beq
      \tilde \rho(\tilde t) ~=~  {\pi \over 2}\, X(\tilde t_0) \, 
                \tilde t_0^{\,5/2} \, \tilde t^{-3/2} ~
\label{res1}
\eeq
where the {\it time-independent}\/ coefficient $X(\tilde t_0)$ is given by
\beq
     X(\tilde t_0) ~\equiv~ 
    \sum_\lambda \,A_\lambda^2\, \tilde \lambda^3 \, 
     \left\lbrace [J_{5/4}(\tilde \lambda \tilde t_0)]^2 + 
              [J_{-5/4}(\tilde \lambda \tilde t_0)]^2  +  
       \sqrt{2}\, J_{5/4}(\tilde \lambda \tilde t_0) 
           J_{-5/4}(\tilde \lambda \tilde t_0) \right\rbrace~. 
\label{Xdef}
\eeq
This demonstrates that asymptotically, the total 
energy density falls as $\tilde \rho(\tilde t)\sim \tilde t^{-3/2}$
 {\it regardless}\/ of the presence of the excited axion Kaluza-Klein modes.
Moreover, in the ``double-asymptotic'' cases 
in which we also have $\tilde \lambda \tilde t_0\gg 1$ for all 
$\tilde \lambda$, we can further approximate
\beq    X(\tilde t_0) ~\approx~  
    {1\over \pi \tilde t_0} \, \sum_\lambda \tilde \lambda^2 A_\lambda^2 ~=~
    {1\over \pi \tilde t_0} ~,
\label{doubleone}
\eeq
where we have used the identity (\ref{identity}) in the last equality.
We then find
\beq
     \tilde \rho(\tilde t) ~\approx~ {1\over 2} 
         \,\left({\tilde t\over \tilde t_0}\right)^{-3/2}~,
\label{doubletwo}
\eeq
which is consistent with the initial energy density $\tilde \rho(\tilde t_0)=1/2$.
However, for practical purposes, we shall focus on the expressions
(\ref{res1}) and (\ref{Xdef})
in which $\tilde t$ is taken large but $\tilde t_0$ is kept arbitrary.
This is because $\tilde\lambda \tilde t_0$ can occasionally be relatively small 
(particularly for the lightest eigenvalues)
even if $\tilde t_0$ itself is relatively large.  
Finally, in the four-dimensional $y\to\infty$ limit, we find that
$\tilde \lambda_k\to ky$ for $k\geq 1$, while $\tilde \lambda_0\to 1$. 
Thus, in this limit, the excited Kaluza-Klein modes decouple and we 
obtain the four-dimensional result
\beq
       X_{\rm 4D}(\tilde t_0)~=~   
     [J_{5/4}(\tilde t_0)]^2 + 
              [J_{-5/4}(\tilde t_0)]^2  +  
       \sqrt{2}\, J_{5/4}(\tilde t_0) 
           J_{-5/4}(\tilde t_0)~.
\label{X4Ddef}
\eeq
As expected, this approaches the value given in (\ref{doubleone}) for $\tilde t_0\gg 1$.

We see from (\ref{res1})
that although the {\it rate}\/ of energy loss remains fixed at asymptotic times,
the presence of the Kaluza-Klein states can nevertheless change the overall
 {\it value}\/ of the energy.  
This is because the oscillating excited Kaluza-Klein states
can change the rate of energy loss at {\it intermediate}\/ 
times, possibly leading to an enhanced or diminished energy at late times.   
The cumulative effect of these Kaluza-Klein states 
at late times is encoded within the expression $X(\tilde t_0)$.  
Therefore, in order to understand the effect of the Kaluza-Klein states
on the energy dissipation rate, we need to understand the
behavior of $X(\tilde t_0)$ as a function of the radius variable 
$y\equiv (m_{\rm PQ}R)^{-1}$ for a fixed initial time $\tilde t_0$.
 
The results are rather surprising.  Of course, for $\tilde t_0\gg 1$, we are 
in the ``double-asymptotic'' regime $(\tilde t_0,\tilde t)\gg 1$ 
for which (\ref{doubleone}) and (\ref{doubletwo}) are expected to apply.
We therefore find that in such cases the presence of the Kaluza-Klein modes
does {\it not}\/ alter the energy density relative to 
the energy density that would have been obtained in four dimensions.
In other words, even though we have an infinite set of 
Kaluza-Klein axion modes which are induced into oscillation as a result
of the initial displacement of the axion zero-mode,  
these oscillations nevertheless change the time-development of the zero-mode
in a compensatory manner so that the total oscillation energy density 
as a function of time is exactly preserved.  This indicates that 
in such situations, the higher-dimensional axion scenarios are no less
viable than the usual four-dimensional scenarios.

\begin{figure}[ht]
\centerline{
      \epsfxsize 3.6 truein \epsfbox {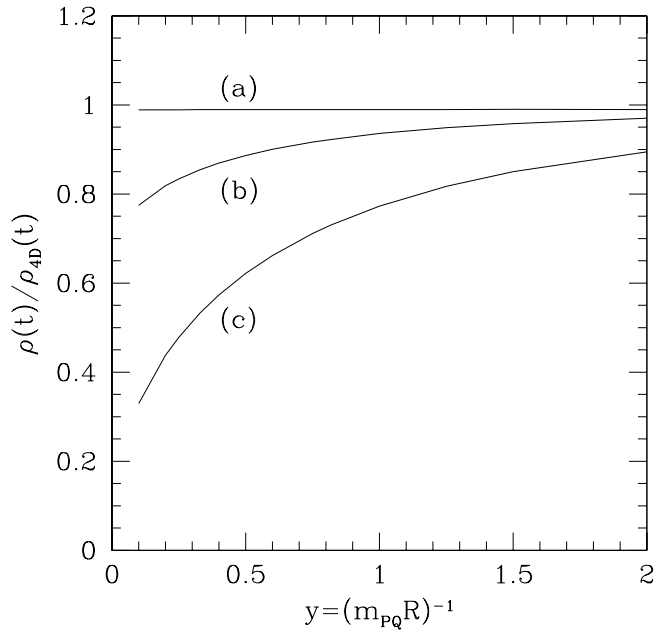}}
\caption{The energy-dissipation ratio factor
      $\rho(t)/\rho_{\rm 4D}(t)= X(\tilde t_0)/X_{\rm 4D}(\tilde t_0)$
      as a function of $y\equiv (m_{\rm PQ}R)^{-1}$, assuming
      (a) $\tilde t_0 =10$; (b) $\tilde t_0=1$; and (c) $\tilde t_0=0.1$.
      All cases reduce to the 
     the usual four-dimensional result 
       as $y\to\infty$, with  the Kaluza-Klein axion states decoupling.
        However, as the size of the extra dimension grows and $y$ decreases 
        from infinity, we see that the net effect of the Kaluza-Klein states
       is to dissipate the relic oscillation energy density more
     {\it rapidly}\/, leading to smaller relic oscillation energy densities at final times.
      The size of this effect depends on $\tilde t_0$, with 
      smaller values of $\tilde t_0$ corresponding to sizable effects
     at large values of $y$, while for larger values of $\tilde t_0$ 
     this effect is delayed until correspondingly smaller values of $y$.   
     In general, this effect becomes substantial for $y \tilde t_0 \lsim {\cal O}(1)$.
     Note that all curves tend to zero  in the $y\to 0$ limit, implying
     an infinitely rapid dissipation of the relic oscillation energy density
     in the full five-dimensional limit.}
\label{relics}
\end{figure}

Even more surprising, however, is the situation that arises for smaller $\tilde t_0$.
In such cases, we cannot use the ``double-asymptotic'' expression (\ref{doubleone}),
and we must resort to the full expression in (\ref{Xdef}).
We then find that $X(\tilde t_0)$ is {\it smaller}\/ than the four-dimensional
value given in (\ref{X4Ddef}), which implies that
\beq
             \rho(t) ~< ~ \rho_{\rm 4D}(t)~.
\eeq
In other words, even though the excited Kaluza-Klein states are triggered
into oscillation by the initial displacement of the zero-mode, these
Kaluza-Klein states dissipate the oscillation 
energy {\it more efficiently}\/ and result in
a net {\it decrease}\/ in the oscillation energy as a function of time
relative to the four-dimensional case.
This situation is shown in Fig.~\ref{relics}, where we plot
$\rho(t)/\rho_{\rm 4D}(t)= X(\tilde t_0)/X_{\rm 4D}(\tilde t_0)$
as a function of $y\equiv (m_{\rm PQ}R)^{-1}$ for three different values
of $\tilde t_0$. 
Note that in the $y\rightarrow 0$, $t_0 \to 0$ limit,
the ratio $\rho(t)/\rho_{4D}(t)$ decreases as $\sqrt{y}$.

We thus conclude that in such situations,
the presence of coupled relic Kaluza-Klein axion oscillations
can actually {\it weaken}\/ the usual four-dimensional upper bounds on  
$\hat f_{\rm PQ}$.  This implies that it may be possible to consider higher
values of $\hat f_{\rm PQ}$ than are usually allowed in four dimensions, thereby 
further diminishing the axion couplings to matter and providing 
yet another higher-dimensional method of achieving an ``invisible'' axion.
We hasten to point out, however, that the size of this effect depends crucially
on the value of $y$ as well as
on the initial time $\tilde t_0$ at which the axion potential is established.
In general, for a given value of $\tilde t_0$, we can expect to see 
a sizable deviation from the four-dimensional asymptotic result only when  
$\lambda \tilde t_0 \lsim {\cal O}(1)$ for the lightest eigenvalues.  
This implies $y\tilde t_0\lsim {\cal O}(1)$.
Taking $t_0=10^{-5}$ seconds (corresponding to the QCD phase transition)
and a reference value $m_{\rm PQ}=10^{-4}$ eV, we find
$\tilde t_0\equiv m_{\rm PQ}t_0 \approx 10^{6}$.
(By contrast, the current cosmological time is given by 
$\tilde t\approx 10^{29}$, corresponding to $t=10^{18}$ seconds.)
This implies that in practice, we should not expect to see a sizable decrease in
the cosmological relic oscillation energy density 
unless $y\lsim {\cal O}(10^{-6})$.
However, this assumes that a particular value of $m_{\rm PQ}$, 
which in turn assumes a particular fixed value of $\hat f_{\rm PQ}$
via (\ref{defs}).  Changing $\hat f_{\rm PQ}$ can therefore change
this result substantially. 
We shall discuss this possibility in Sect.~5.

Finally, let us briefly discuss the issue
of axion lifetimes as they relate to these cosmological relic oscillations. 
Ordinarily, relic axion oscillations 
are important precisely because the usual four-dimensional axion is so 
long-lived.  However, in higher dimensions, we have seen in Sect.~3
that the heavy Kaluza-Klein axion modes
become more and more unstable, and hence cannot be expected to
survive over the long cosmological time scales we have been assuming.
It is therefore natural to wonder whether this places our conclusions
about enhanced energy dissipation rates in jeopardy.
In other words, it is not {\it a priori}\/ obvious that we are
justified in neglecting the excited Kaluza-Klein decay widths in
(\ref{newdiffeq}). 
However, we have seen that relic axion oscillations
in higher dimensions are primarily sensitive not to the heaviest  
axion modes, but to the lightest modes.  For sufficiently large radii,
these lightest modes have masses which are relatively close
to the mass of the axion zero-mode, and we have already seen in Fig.~\ref{massfig}
that this in turn is bounded from above by
the usual four-dimensional axion mass $m_{\rm PQ}$.  Thus, the
lightest axion Kaluza-Klein modes continue to be extremely long-lived, 
and will therefore survive to induce the enhancement of the 
oscillation energy dissipation rate that we have observed.

\section{Some numbers, bounds, and constraints}

Let us now combine our different results from the preceding
sections.  Our goal will be to determine the extent to which
a self-consistent picture of axion energy scales emerges from the
previous results.

We begin with the three fundamental equations given in 
(\ref{fhatdef}) and (\ref{defs}).
For simplicity, in (\ref{defs})
we shall take $g^2=g_{\rm GUT}^2 = 1/2$,
and in (\ref{fhatdef}) we shall take $\delta=1$.
We can therefore combine these three equations in order to 
express $y$ in terms of the fixed
quantities $\Lambda_{\rm QCD}$, $M_{\rm string}$ (the fundamental underlying
mass scale in the theory), and the effective axion decay constant
$\hat f_{\rm PQ}$ (our measure of ``invisibility'').  
For simplicity we shall also take $f_{\rm PQ}=M_{\rm string}$, 
since we want to have only one fundamental mass scale in the problem.  
This then yields the result
\beq
         y  ~\approx~   {16 \pi^2 \over \xi}  \,   
      { M_{\rm string}^3 \over \hat f_{\rm PQ}  \Lambda_{\rm QCD}^2 }~.
\label{firstresult}
\eeq

We know that $M_{\rm string}$ cannot be lower than $\approx {\cal O}(1~{\rm TeV})$, 
and $\Lambda_{\rm QCD}$ is absolutely fixed ($\approx 250$ MeV).  
We therefore find that in general, $y$ is bounded from below: 
\beq      y~\gsim~ y_{\rm min} ~\approx~  { 2.5 \times  10^{12} ~{\rm GeV}\over \xi \, 
                 \hat f_{\rm PQ}}~.
\label{ybound}
\eeq
Thus, assuming $\xi\approx {\cal O}(1)$, we see that
$\hat f_{\rm PQ}\approx 10^{12}$ GeV is consistent with
having $y\approx {\cal O}(1)$.  Remarkably,  this is precisely the region 
where we expect to find the axion mass becoming independent of
$\hat f_{\rm PQ}$, as shown in Fig.~\ref{massfig}.

While this shows the self-consistency of the situation depicted in
Fig.~\ref{massfig}, a natural question arises as to whether it is
possible to tolerate larger values of $M_{\rm string}$. 
Indeed, slightly larger values of $M_{\rm string}$ [\eg, in the
${\cal O}(10~{\rm TeV})$ range] may be preferred on the basis
of detailed comparisons with experimental data. 
Ordinarily, it might seem to be impossible to increase the value of $M_{\rm string}$
any further, because we see from (\ref{firstresult}) that increasing $M_{\rm string}$
requires increasing $\hat f_{\rm PQ}$ in order to maintain the ${\cal O}(1)$
values of $y$ (as preferred on the basis of Fig.~\ref{massfig}),  
and increasing $\hat f_{\rm PQ}$ generally
runs into difficulties with cosmological relic oscillation 
energy densities overclosing the universe.  
However, we have seen in Sect.~4 that the Kaluza-Klein axion modes   
may be capable of dissipating this excess energy density more rapidly so as to evade
these bounds.  The question then arises:  to what extent can we increase
$\hat f_{\rm PQ}$, thereby making the axion increasingly ``invisible'',
without disturbing the relic energy density bounds?

Note that increasing $\hat f_{\rm PQ}$ has a number of effects.
First, as $\hat f_{\rm PQ}$ increases,
we find from (\ref{ybound}) that $y_{\rm min}$ decreases.
This means that $y$ can be chosen even smaller. 
However, we also find from (\ref{defs}) that $m_{\rm PQ}$ decreases,
which in turn implies that $\tilde t_0\equiv m_{\rm PQ} t_0$ 
(the dimensionless time of the QCD phase transition) also decreases.
Defining 
$y_{\rm crit}$ to be the critical value of $y$
at which we start to observe a significant decrease in the relic
oscillation energy density, 
we have already seen in Sect.~4 that $y_{\rm crit} \approx \tilde t_0^{-1}$.
Thus, as $\tilde t_0$ decreases, we see that
$y_{\rm crit}$ increases, implying
that it becomes easier to compensate for the effect of having increased
$\hat f_{\rm PQ}$ in the first place.
Indeed, this suggests that there might be 
an alternative, self-consistent, significantly higher value 
of $\hat f_{\rm PQ}$ than previously thought.

In order to determine this self-consistent value of $\hat f_{\rm PQ}$,
we first note from (\ref{defs}) that
\beq
           m_{\rm PQ} ~\approx~ {
       (2.5 \times 10^{-3} ~ {\rm GeV}^2)\,\xi \over \hat f_{\rm PQ}}~.
\label{fivethree}
\eeq
This in turn implies that 
\beq
          \tilde t_0 ~\equiv~ m_{\rm PQ}\,t_0 ~\approx~ 
          {(3.8 \times 10^{16}~{\rm GeV})\, \xi \over \hat f_{\rm PQ}}
\eeq
where we have taken $t_0\approx 10^{-5}$ seconds 
(corresponding to the QCD phase transition).
We thus have 
\beq
              \xi \,y_{\rm crit}~\approx~ 
          {\hat f_{\rm PQ} \over 3.8 \times 10^{16}~{\rm GeV}}~,
\label{intermed}
\eeq
which implies that we can obtain $y_{\rm crit}\approx {\cal O}(1)$ 
simply by taking $\xi\approx {\cal O}(1)$ and 
\beq
               \hat f_{\rm PQ}~\approx~ 3.8\times 10^{16}~{\rm GeV}~\approx~ 
               M_{\rm GUT}~.
\eeq
In other words, if we take $\hat f_{\rm PQ}\approx M_{\rm GUT}$,
then we have a self-consistent solution with
$y_{\rm min}\approx {\cal O}(1)$, $y_{\rm crit}\approx {\cal O}(1)$, 
and $M_{\rm string}\approx 20$ TeV.
For such values, the axion mass is independent of $\hat f_{\rm PQ}$,
and the Kaluza-Klein modes begin to induce a significant reduction
in the final relic oscillation energy density which can in principle compensate
for the increase in $\hat f_{\rm PQ}$.  
Remarkably, this analysis suggests that 
$\hat f_{\rm PQ}$, the effective 
Peccei-Quinn symmetry-breaking scale, may be related to 
$M_{\rm GUT}$, the effective four-dimensional GUT symmetry-breaking scale.

Of course, there is still one constraint that we have not imposed in the
above analysis:  we have not restricted the size of the radius $R$ of the
extra spacetime dimension.  In principle, this is not a problem
because the Standard-Model fields are restricted to a D-brane, and thus there
are no bounds on the sizes of such transverse extra dimensions that can 
arise from Standard-Model processes.
However, gravity is generally free to propagate into whatever extra dimensions
exist, leading to the additional constraint
$R\lsim {\cal O}$(millimeter).  It is therefore important to understand how this
additional constraint limits the above scenarios.  
We stress, however, that imposing this additional constraint
relies on the assumption that gravity is indeed free to propagate 
in the extra dimensions.  Recent
ideas concerning ``gravity localization''~\cite{RS} 
have shown that this need not always be the case.

If we do assume this to be the case, however, then 
the above scenarios are significantly restricted.
Requiring $R^{-1}\gsim 10^{-4}$~eV implies\footnote{
       Note that imposing this constraint is actually somewhat 
       subtle, and depends on a choice of which variables to
       hold fixed.  In the analysis in this
       section, we have been taking $\hat f_{\rm PQ}$ and $y$ 
       as inputs, and treating $M_{\rm string}$, $R$, and $m_{\rm PQ}$
       as derived quantities.  Thus, with this convention,
       $\hat f_{\rm PQ}$ is considered to be independent of $R$,
       while $f_{\rm PQ}$ (identified with $M_{\rm string}$)
       is considered to be an $R$-dependent quantity.
       Note that this procedure exactly mimics the situations~\cite{DDG,ADD}
       in which the GUT and Planck scales are lowered by extra 
       spacetime dimensions:
       it is always the ``measured'', large, four-dimensional scale that
       is held fixed, while the reduced higher-dimensional scale is
       viewed as a function of $R$.  We therefore continue this
       convention in the present case even though neither $\hat f_{\rm PQ}$
       nor $f_{\rm PQ}$ has been experimentally measured.}
that 
\beq
         m_{\rm PQ}~\gsim~ {10^{-13}~{\rm GeV}\over y}~.
\eeq
Moreover, using (\ref{fivethree}), we find that this implies that
\beq
        \hat f_{\rm PQ} ~\lsim~ (2.5\times 10^{10}~{\rm GeV}) ~ \xi y~.
\label{uppboundf}
\eeq
This therefore sets an {\it upper bound}\/ on $\hat f_{\rm PQ}$ as a function
of $y$. 
Combining this upper bound with (\ref{firstresult}), we thereby obtain the
constraint 
\beq
            \xi y ~\gsim~ 10 \left( {M_{\rm string}\over 
                  1\, {\rm TeV}}\right)^{3/2}~.
\eeq
Thus, the lower limit for $y$ depends crucially on the model-dependent
parameter $\xi$ and the value we choose for $M_{\rm string}$.
Since $\xi$ reflects the PQ charges of the ordinary fermions,
it is not unreasonable to assume that $\xi$ may be somewhat larger than 1.
Taking $M_{\rm string}\approx 1$~TeV therefore still enables 
us to have $y\approx {\cal O}(1)$.
However, despite this fact, we can combine 
(\ref{intermed}) with (\ref{uppboundf}) to show that 
\beq
        y_{\rm crit} ~\lsim~ (6.6\times 10^{-7})\, y~.
\eeq
Thus the value of $y$ in this case is always significantly larger 
than the critical value that would be required in order to reduce
the relic oscillation energy density below its usual four-dimensional value.

Of course, this does not disturb the self-consistency of this scenario.
As a result of (\ref{uppboundf}), $\hat f_{\rm PQ}$ may
still be in the range that satisfies the usual four-dimensional bounds,
and we have seen in Sect.~4 that the presence of the Kaluza-Klein axion states 
does not increase the final relic oscillation energy 
density relative to the four-dimensional case.
Moreover, these axions continue to be virtually ``invisible''
against direct detection and/or subsequent interactions 
as a result of the decoherence effect discussed in Sect.~3.     
Thus this picture continues to be self-consistent, and continues to lead to
an invisible axion.
Furthermore, although we have restricted our analysis to the case
of a single extra dimension, the corresponding constraints in higher
dimensions may be significantly weaker.
In any case, a much more detailed analysis is necessary in order to 
make the above numerical bounds more precise,
and to determine whether 
further experimental constraints may be imposed.

There are also other phenomenological constraints that may be imposed,
particularly constraints that are insensitive to axion time-evolution.
Good examples of this would be axion missing-energy signatures,
such as might arise from the decay $K^+\to \pi^+ a'$.
In four dimensions, this process has a branching ratio
which scales as $\hat f_{\rm PQ}^{-2}$, leading to a bound 
$\hat f_{\rm PQ}\gsim 10^4$~GeV.
In five dimensions, by contrast, this branching ratio scales as
\beq
       {\rm BR}(K^+\to \pi^+ a') ~\sim~ 
       {1\over \hat f_{\rm PQ}^2} \, \sum_{n=0}^{Rm_K} 1 ~\sim~
         {R m_K\over \hat f_{\rm PQ}^2 } ~\sim~ 
             \left( {m_K\over M_{\rm string}}\right) {1\over f_{\rm PQ}^2}~.
\eeq
For $m_K\approx 500$~MeV,
this then implies the constraint 
\beq
         f_{\rm PQ} ~\gsim~ \sqrt{{ m_K\over M_{\rm string}}}\, 10^4~{\rm GeV}~.
\eeq
Taking $f_{\rm PQ}=M_{\rm string}$ then leads to the 
bound $M_{\rm string}\gsim 370$~GeV,
which is consistent with the idea of lowering the string scale 
to the TeV-range.
Of course, we reiterate that a much more detailed analysis 
of these and other processes
is necessary in order to sharpen these bounds and constraints. 

Finally, it may also happen that an axion process  
in higher dimensions exactly reproduces the four-dimensional result.
As an example of this, let us consider  
an $a'$-exchange process at zero momentum transfer
(\eg, $F\tilde F\to a'\to F\tilde F$).
The amplitude for such a process is given by
\beq
         A  ~=~ {1\over f_{\rm PQ}^2} \, \langle a' a'\rangle ~\sim~
         {1\over \hat f_{\rm PQ}^2} \, \sum_{m,n=0}^\infty 
           r_m r_n \langle a_m a_n\rangle~
\eeq
where $\langle B A\rangle$ denotes the propagator from state $B$ to state $A$.
Passing to the mass-eigenstate basis $\hat a_\lambda$ via (\ref{masseig})
and using the zero-momentum propagator 
$\langle \hat a_{\lambda'} \hat a_\lambda\rangle= 
\delta_{\lambda' \lambda}/\lambda^2$,
we find that this amplitude then takes the form
\beq
       A ~=~ {1\over \hat f_{\rm PQ}^2 m_{\rm PQ}^2}\, 
        \sum_\lambda  \,{1\over \tilde \lambda^2}\,  
            \sum_{m,n=0}^\infty \,r_m r_n \,U_{\lambda m} U_{\lambda n} 
        ~=~ {1\over \hat f_{\rm PQ}^2 m_{\rm PQ}^2}~.
\eeq
Note that we have used (\ref{otheridentity}) followed by (\ref{identity})
in the final equality.
However, we see that the final result is nothing but the amplitude that we would
have obtained in four dimensions for an axion that couples with the 
usual four-dimensional coupling
$\hat f_{\rm PQ}$ and has the usual four-dimensional mass $m_{\rm PQ}$.
Thus, in this case, we obtain no new bounds coming from
such $a'$-mediated processes.  This stands in stark contrast to the analogous
case of graviton-mediated processes, from which one can generally
derive stringent bounds on the radii of the extra spacetime dimensions.
Of course, we stress that this result holds only for zero momentum
transfer, and is likely to be different when sizable momenta are
carried by the intermediate axion state.

\section{The Standard-Model dilaton}

The above considerations about placing a Standard-Model singlet field in the 
bulk are actually quite general, and transcend the specific example of the QCD
axion.  To illustrate this point, let us briefly consider the case of another 
conjectured particle, the Standard-Model dilaton. This particle is introduced 
into the Standard Model in order to restore the classical scale invariance broken
by mass terms. The scale-invariant extension of the scalar sector of 
Standard Model is given by~\cite{coleman}
\beq
    {\cal L} ~=~ \half (\partial_\mu D)^2 +(D_\mu \phi)(D^\mu\phi)^\dagger
              -V_0(\phi,\sigma)
\eeq
where $\phi$ is the Higgs field and where the Standard-Model dilaton field 
$D$, like the axion field $a$, is
written in terms of a decay constant $f_D$ via a relation of the form
\beq
               D~=~ f_D \, \exp(\sigma/f_D)~.
\eeq
By suitably choosing the parameters in the tree-level Higgs potential
$V_0(\phi,\sigma)$, we can arrange $\langle \sigma\rangle =0$.   Consequently
$\langle D\rangle = f_D$ represents the mass scale at which dilatation 
invariance is spontaneously broken. 
After quantum corrections are included, the scalar potential can be written 
in the form~\cite{bd}
\beq
     V ~=~ {D^4\over f_D^4} {\cal V} ~~~~~~{\rm where}~~~
      {\cal V}~\equiv~ 
  \left[ V_0(\phi,0) +V^{(1)}(\phi,0)-\Delta(\phi,0) \ln{D\over f_D}\right]~.
\label{dilpot}
\eeq
Here $V^{(1)}(\phi,0)$ is the one-loop contribution to the effective 
potential and  $\Delta(\phi,0)$ is the divergence of the dilatation current.
The presence of $\Delta(\phi,0)$ in this expression breaks
the scale invariance, and gives rise to a dilaton mass
\beq
         m_D^2~=~ -{4\langle\Delta\rangle \over f_D^2}
\eeq
where $\langle {\cal V}\rangle = \Delta/4$
at the minimum of the potential (\ref{dilpot}).
Note that in the Standard Model, the heavy top-quark mass leads to
$\langle\Delta\rangle > 0$.  Thus, in order to change the sign of 
$\langle\Delta\rangle$, one requires additional
heavy Higgs-boson contributions for the stability of the dilaton potential.

Let us now consider what happens when the dilaton field propagates in a 
five-dimensional bulk
and therefore has a Kaluza-Klein decomposition of the form
\beq
     D(x^\mu,y) ~=~ \sum_{n=0}^\infty \, D_n(x^\mu) \cos\left({ny\over R}\right)~.
\label{dilKK}
\eeq
The five-dimensional action for the dilaton then takes the form
\beq
  {\cal S} ~=~ \int d^4x \,dy\, M_s\,\left[ \half (\partial_M D)^2 - V(x) \delta(y) \right]
\label{dillag}
\eeq
where $V$ is given in (\ref{dilpot}). 
Substituting (\ref{dilKK}) into 
(\ref{dillag}) and integrating over the fifth dimension then gives rise to the
effective four-dimensional Lagrangian
\beqn
       {\cal L}_{\rm eff} &=& \half\sum_{n=0}^\infty
      (\partial_\mu D_n)^2 - \half \sum_{n=1}^\infty {n^2\over R^2} D_n^2
      - {1\over {\hat f}_D^4} \left(\sum_{n=0}^\infty r_n D_n\right)^4  \times\nonumber\\
       && ~~~~~~~~ \times \left[V_0(\phi,0) + V^{(1)}(\phi,0) - \Delta(\phi,0)
       \ln \left( {1\over {\hat f}_D}\sum_{n=0}^\infty r_n D_n\right)\right]
\eeqn
where ${\hat f}_D \equiv\sqrt{2\pi R M_{\rm string}} f_D$ and where we have canonically 
normalized the dilaton kinetic terms. 
The minimum of the effective dilaton potential 
therefore occurs at
\beqn
      \langle D_0\rangle &=&  {\hat f}_D\nonumber\\
      \langle D_n\rangle &=& 0   ~~~{\rm for~all}~ n\geq 1~.
\eeqn

We can derive the dilaton mass matrix by considering 
the local curvature of the effective dilaton potential
near its minimum.  This is 
completely analogous to the axion case (\ref{massderivatives}),
and gives rise to
\beq
     {({\cal M}_D)}^2_{nn'} ~\equiv~ {n^2\over R^2} \delta_{nn'} ~-~ 
      {4\Delta\over {\hat f_D}^2 } \, r_n r_{n'}~.
\eeq
Remarkably, this mass matrix has exactly the same structure as in the axion 
case (\ref{matrixalg}),
and consequently the physical implications will be identical to those for
the axion.
Of course, this result is expected since the dilaton and axion are both
Nambu-Goldstone bosons of a spontaneously broken symmetry, and consequently
have similar couplings to the anomalous divergences of their respective currents.

\section{Conclusions}

In this paper, we have studied 
some of the novel effects that arise when the QCD axion
is free to propagate in the bulk of large extra spacetime dimensions.
First, we found that under certain circumstances, the mass of the axion 
can become independent of the energy scale associated with the breaking of 
the Peccei-Quinn symmetry.  Because this energy scale determines the
couplings between the axion and ordinary matter,
we now have the freedom to adjust the strength of these couplings
without disturbing the axion mass. 
This can therefore provide a new mechanism for achieving
an invisible axion.

Second, we pointed out that in such higher-dimensional scenarios,
the axion will typically experience {\it laboratory axion oscillations}\/ 
which are completely analogous to neutrino oscillations.
This is therefore a new and unexpected phenomenological feature
for axions that does not exist in the usual four-dimensional case.
Moreover, we found that these laboratory oscillations can cause
axions to ``decohere'' extremely rapidly.
This is therefore a second higher-dimensional phenomenon that may contribute
to an invisible axion.
Moreover, this phenomenon arises for all non-zero radii.

Third, we discussed the role that excited Kaluza-Klein axion states may
play in axion-mediated processes and axion decays.  This enabled us
to propose several direct experimental tests of the proposed higher-dimensional
nature of the axion.

Finally, we found that under certain circumstances, the presence of these
Kaluza-Klein axion modes can significantly accelerate the dissipation of
the energy associated with cosmological relic axion oscillations. 
Moreover, even when these circumstances are not met, we found that
the Kaluza-Klein states do not induce a violation of the usual four-dimensional
bounds.
This demonstrates that such higher-dimensional axion scenarios are
no less viable than their four-dimensional counterparts, and indeed
may even be preferred on the basis of their remarkable ``invisibility''
decoherence properties.

Of course, there are many aspects of higher-dimensional 
axion phenomenology which we have
 {\it not}\/ examined in this paper.  These include 
the role that axions play in stellar evolution,
the thermal production of axions, axionic string decay,
and isocurvature axion fluctuations.
While some of these topics have been discussed in 
Ref.~\cite{CTY},
it will be interesting to further explore the role that 
extra spacetime dimensions can play in these areas.

Although we have focused primarily on the case of QCD axions,
we stress that much of our analysis is completely general and
may apply for other bulk fields as well.  This was explicitly
illustrated in Sect.~6, where we considered the case of the
Standard-Model dilaton.  Similarly, we expect that our analysis
will also apply to other bulk fields such as 
Kaluza-Klein gravitons, string-theoretic dilatons, 
and other bulk moduli.
Indeed, the twin properties of laboratory oscillations and decoherence
leading to ``invisibility'' are likely to play an important
role in experimental searches for such particles, and likewise
an analysis of the effects of their excited Kaluza-Klein modes 
on cosmological evolution is likely to parallel the analysis in Sect.~4. 
Moreover, both of these effects are likely to play an important
role in the all-important questions of dilaton and radion stabilization. 
We therefore leave these issues for future investigation.

\bigskip
\medskip
\leftline{\large\bf Acknowledgments}
\medskip

We wish to thank K.~Orginos, I.~Sarcevic,
and especially J.~Mourad 
for useful discussions.  KRD and TG also wish to acknowledge the hospitality
of the Aspen Center for Physics where part of this work was done.

\vfill\eject
\bigskip
\medskip

\bibliographystyle{unsrt}

\end{document}